\documentclass[]{aastex631}

\usepackage{amsmath}
\usepackage{math}
\usepackage{amssymb}
\usepackage{amsfonts}
\usepackage{amsthm}

\let\oldFootnote\footnote
\newcommand\nextToken\relax

\renewcommand\footnote[1]{%
    \oldFootnote{#1}\futurelet\nextToken\isFootnote}

\newcommand\isFootnote{%
    \ifx\footnote\nextToken\textsuperscript{,}\fi}

\usepackage{stmaryrd}

\usepackage{etoolbox}
\makeatletter
\patchcmd\linenumberpar{\@LN@parpgbrk}{\penalty\@LN@parpgpen\relax}{}{}
\makeatother

\usepackage{labelschanged}
\usepackage[inline]{enumitem}
\setlist[enumerate,1]{label=(\roman*)}
\tabletypesize{\scriptsize}
\usepackage{tikz}
\usepackage{tikzscale}
\usepackage{tikz-3dplot}
\usepackage{pgfplots}
\usetikzlibrary{shapes}
\usetikzlibrary{shadings}
\usetikzlibrary{positioning}
\usetikzlibrary{arrows, arrows.meta}
\usetikzlibrary{calc,3d,shapes, pgfplots.external}
\pgfplotsset{compat=newest}

\usepackage[caption=false]{subfig}

\usepackage{ifthen}

\graphicspath{{./}{./picts/}} 

\makeatletter
\def\input@path{{./picts/}}
\makeatother

\makeatletter
\newcommand{\gettikzxy}[3]{%
  \tikz@scan@one@point\pgfutil@firstofone#1\relax
  \edef#2{\the\pgf@x}%
  \edef#3{\the\pgf@y}%
}
\makeatother

\def\overleafhome{/tmp}

\makeatletter
\begingroup\endlinechar=-1\relax
       \everyeof{\noexpand}%
       \edef\x{\endgroup\def\noexpand\homepath{%
         \@@input|"kpsewhich --var-value=HOME" }}\x
\makeatother

\keywords{Analytical mathematics (38), Astrophysical magnetism (102), Stellar magnetic fields (1610), Solar magnetic fields (1503), Solar magnetic reconnection (1504), Magnetohydrodynamics (1964)}

\begin{document}
\title{Disentangling the Entangled Linkages of Relative Magnetic Helicity}
\shorttitle{Disentangling the Entangled Linkages of Relative Magnetic
  Helicity} \author[0000-0003-1522-4632]{Peter W. Schuck}
\affiliation{Heliophysics Science Division, NASA Goddard Space Flight Center,
  8800 Greenbelt Rd., Greenbelt, MD 20771, USA}
\author[0000-0002-4459-7510]{Mark G. Linton} \affiliation{United States Naval
  Research Laboratory, 4555 Overlook Ave SW, Washington, DC 20375, USA}
\shortauthors{Schuck and Linton} \received{July 20, 2023} \accepted{August 17,
  2023}
\journalinfo{
  This is the Accepted Manuscript version of an article
  accepted for publication in the Astrophysical Journal.  IOP Publishing Ltd
  is not responsible for any errors or omission in this version of the
  manuscript of any version derived from it. This Accepted Manuscript is
  published under a CC BY license.}
\begin{abstract}
Magnetic helicity, $H$, measures magnetic linkages in a volume. The early
theoretical development of helicity focused on magnetically closed systems in
$\Vol$ bounded by $\Surf$. For magnetically closed systems,
$\Vol\in\Rthree=\Vol+\VolC$, no magnetic flux threads the boundary,
$\nhat\cdot\B|_\Surf=0$. \cite{Berger1984a} and \cite{Finn1985} extended the
definition of helicity to relative helicity, $\Helicity$, for magnetically
open systems where magnetic flux may thread the boundary.
\cite{Berger1999,Berger2003} expressed this relative helicity as two gauge
invariant terms that describe the self helicity of magnetic field that closes
inside $\Vol$ and the mutual helicity between the magnetic field that threads
the boundary $\Surf$ and the magnetic field that closes inside $\Vol$. The
total magnetic field that permeates $\Vol$ entangles magnetic fields that are
produced by current sources $\JM$ in $\Vol$ with magnetic fields that are
produced by current sources $\JME$ in $\VolC$. Building on this fact, we
extend \citeauthor{Berger2003}'s expressions for relative magnetic helicity to
eight gauge invariant quantities that simultaneously characterize
both of these self and mutual helicities and attribute their origins to
currents $\JM$ in $\Vol$ and/or $\JME$ in $\VolC$, thereby disentangling the
domain of origin for these entangled linkages. We arrange these eight
terms into novel expressions for internal and external helicity (self) and
internal-external helicity (mutual) based on their domain of origin.  The
implications of these linkages for interpreting magnetic energy is discussed
and new boundary observables are proposed for tracking the evolution of the
field that threads the boundary.
\end{abstract}

\section{Introduction\label{sec:Intro}}
Magnetic helicity is an important astrophysical quantity for understanding
dynamos \cite[]{Moffatt1978}, the emergence of large scale magnetic fields in
the primodial universe \cite[]{Field2000,Brandenburg2006}, galactic jets
\cite[]{Koenigl1985}, the structure of stars
\cite[]{Schrijver2000,Brandenburg2020}, stellar eruptive phenomena
\cite[]{Berger1984b}, and coronal heating \cite[]{Heyvaerts1984}. The concept
of helicity has its mathematical origins in linkages with \cite{GaussWerkeV},
\cite{Calugareanu1959}, and \cite{White1969} and vortex motion with
\cite{Thomson1868}.  There have been five major developments in understanding
magnetic helicity. First, \cite{Woltjer1958} proved that magnetic helicity is
preserved in ideal magnetically closed plasma systems and that a linear
force-free magnetic configuration represents the absolute minimum energy state
for a magnetically closed plasma with a prescribed magnetic
helicity. Second, \cite{Taylor1974,Taylor1986} conjectured that magnetic
helicity was preserved under turbulent reconnection, thus providing a pathway
for plasma to relax to a linear force-free \citeauthor{Woltjer1958}
state. Third, \cite{Frisch1975} demonstrated that helicity can inverse cascade
in the spectral domain to the largest scales accessible to the system,
producing large scale magnetic fields. Fourth, \cite{Berger1984a} and
\cite{Finn1985} extended the definition of magnetic helicity to magnetically
open systems by introducing a reference magnetic field that matches the `open'
flux threading the boundary surface $\Surf$ of the volume of interest
$\Vol$\----the so-called ``relative magnetic helicity.''  Fifth,
\cite{Berger1984a} also showed that the evolution of this relative magnetic
helicity for an ideal plasma could be determined from boundary
observables. Further refinements on these five major developments have since
been made. \cite{Berger1984b} adapted \citeauthor{Taylor1974}'s conjecture to
the relative helicity of open systems, arguing that the relative helicity is
preserved during solar flares. \cite{Berger1999,Berger2003} later partitioned
the relative magnetic helicity into two further gauge invariant topological
quantities: the `self' helicity representing the linkages of the magnetic
field that closes in $\Vol$ that we term the ``closed-closed
  helicity'' and the `mutual' helicity representing the linkages between the
open magnetic field that threads the boundary and the magnetic field that
closes inside $\Vol$ that we term the ``open-closed helicity.'' We
  have modified this terminology because `self' and/or `mutual' helicity have
  a variety of meanings in the literature in terms of isolated flux tubes
  \cite[]{Berger1984a,Berger1984b,Berger1986,Demoulin2006}, relative helicity
  of distributed fields in a volume $\Vol$ \cite[]{Berger1999,Berger2003},
  relative helicity in multiple disjoint subdomains $\Vol=\cup_{i=1}^N\Vol_i$
  \cite[]{Longcope2008}, winding helicity in subdomains
  \cite[]{Candelaresi2021}, etc.  Recently, \cite{Schuck2019} recast the
helicity transport across the boundary in \cite{Berger1984a} in a
manifestly gauge invariant way and proved that the instantaneous time rate of
change of relative helicity was independent of the instantaneous time rate of
change of the flux threading the boundary $\Surf$.  \par
The magnetic helicity is
\begin{subequations}
\begin{equation}
H\equiv\int\limits_{\Vol}{d^3x}\,\AV\cdot\B=\int\limits_{\Vol}{d^3x}\,\AV\cdot\grad\cross\AV,
\end{equation}
where $\AV$ is the vector potential and
\begin{equation}
\B=\grad\cross\AV,\label{eqn:B}
\end{equation}  
\end{subequations}
is the magnetic field.  Magnetic helicity is challenging to quantify because
because $\AV$ itself is not directly observable and thus there is gauge
freedom in specifying the vector potential $\AV$ that determines $\B$
through Equation~(\ref{eqn:B}). Thus, under a local gauge
transformation\footnote{The local gauge symmetry of Maxwell's equations
  implies a conserved \citeauthor{Noether1918b} current by Emma
  \citeauthor{Noether1918b}'s second theorem
  (\citeyear{Noether1918b}). However, these currents do not generally
  correspond to physical observables as these currents are not themselves
  gauge invariant \cite[]{Karatas1990}.\label{ft:Noether}}
$\AV\rightarrow\AV+\grad\gauge$, the magnetic field remains unchanged, but the
helicity becomes \cite[see for example][]{Schuck2019}
\begin{equation}
H\rightarrow{H}-\oint\limits_{\Surf}{dS}\,\nhat\cdot\left(\gauge\,\B\right),\label{eqn:Gauge}
\end{equation}
where $\nhat$ is the normal pointing into $\Vol$ on $\Surf$.\par
\begin{figure}
  \centering
  \newlength{\mywidth}
  \setlength{\mywidth}{1.25in}
    \centering
    \subfloat[The magnetic field: $\B$]{\includegraphics[width=\mywidth,angle=90]{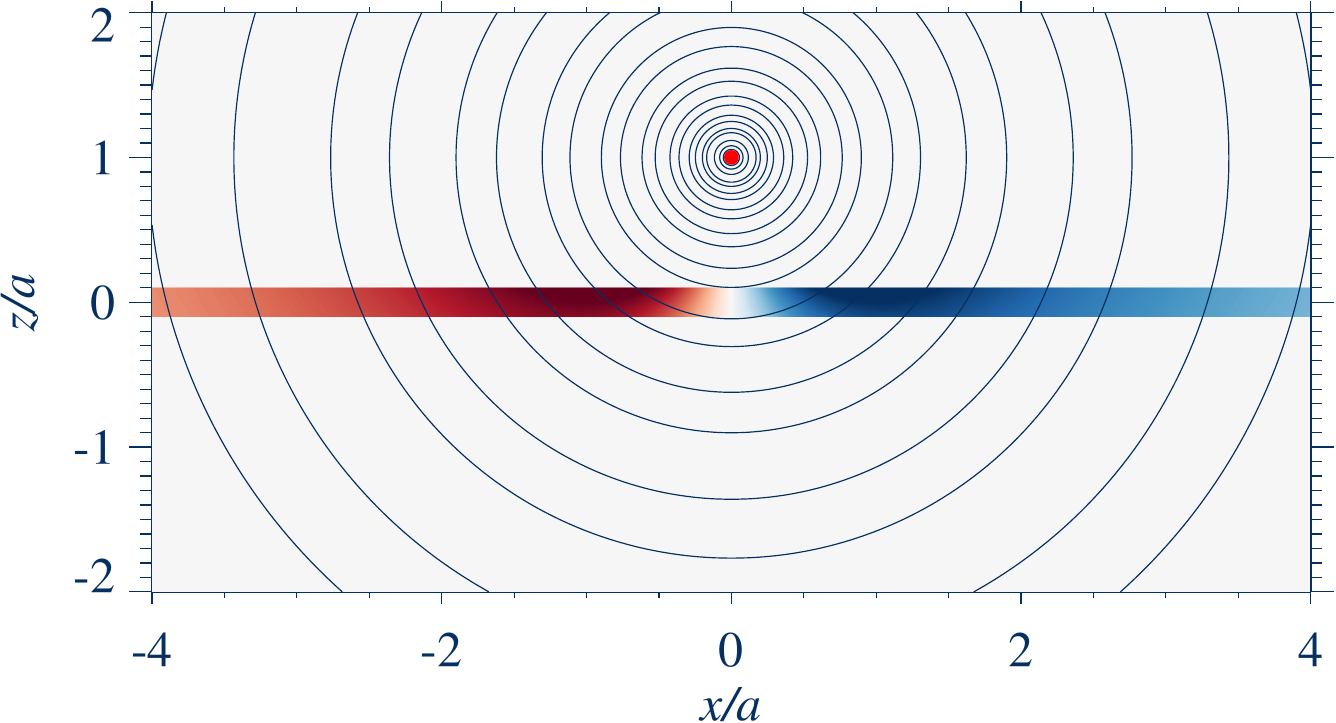}
      \label{fig:B}}
    \subfloat[The ``potential'' field: $\BPot$]{
    \includegraphics[width=\mywidth,angle=90,viewport=0 0 354 580 ,clip=]{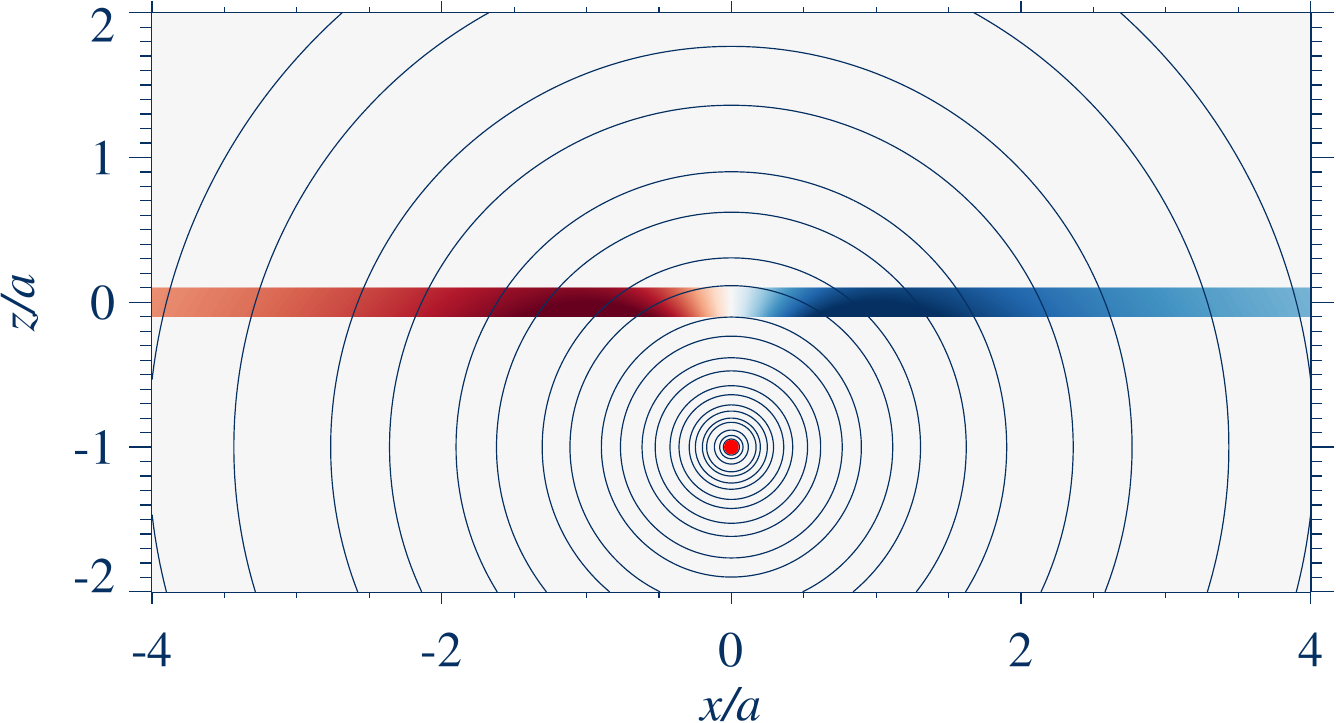}
    \label{fig:BP}}
    \subfloat[The ``closed'' field: $\Bcl$]{
    \includegraphics[width=\mywidth,angle=90,viewport=0 0 354 580 ,clip=]{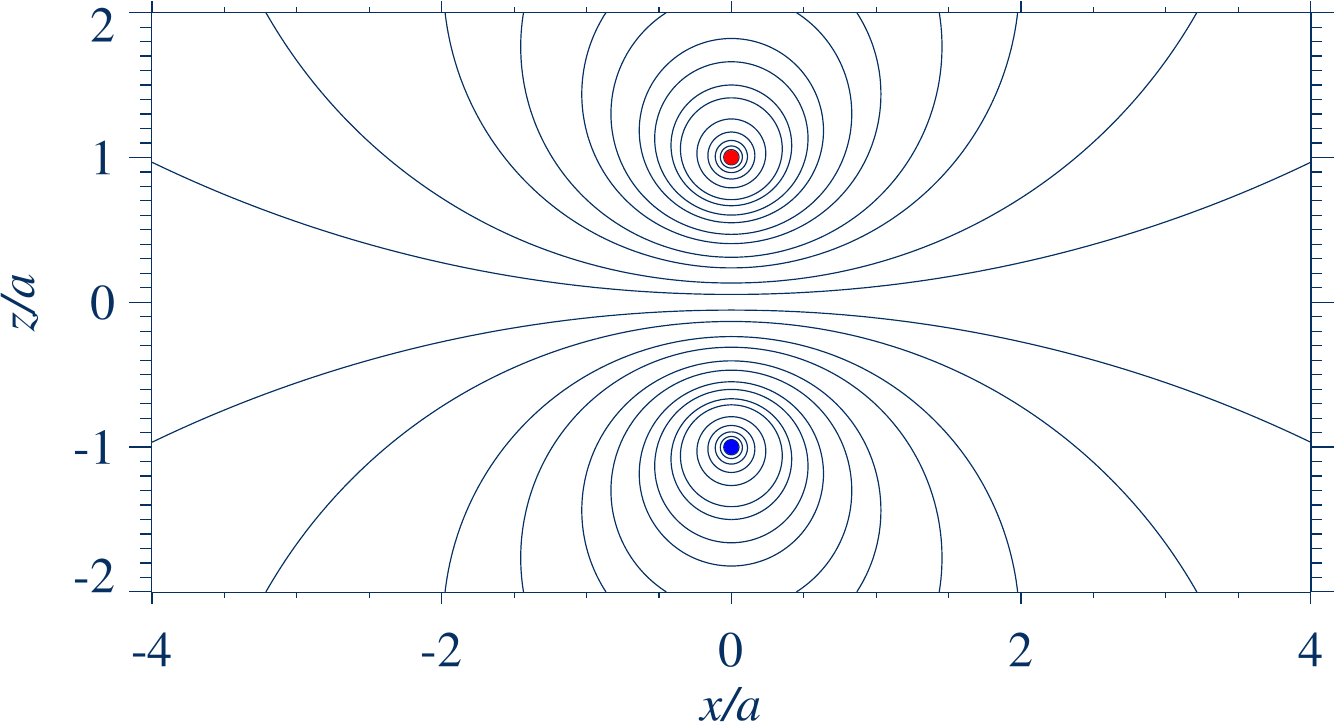}
    \label{fig:BC}}
    \caption{The entangled physical origins of \subref{fig:B} the
      magnetic field $\B=\BPot+\Bcl$ for $z>0$ when it is decomposed
      into the fields \subref{fig:BP} $\BPot$ which thread the
      boundary at $z=0$ and are potential for $z>0$ and
      \subref{fig:BC} $\Bcl$ which close on themselves for $z>0$. A
      red dot indicates a line current, $I\,\yhat$, directed away from the observer
      and a blue dot indicates a line current, $-I\,\yhat$, towards the
      observer. The black lines are contours of the vector potential
      that trace magnetic field lines. The color scale along $z=0$
      corresponds to the vertical magnetic field component with
      red/blue corresponding to up/down.  All three magnetic fields,
      $\B$, $\BPot$, and $\Bcl$, are \emph{produced} by a physical
      current $I\,\yhat$ at $x/a=0; z/a=1$, but $\BPot$ is \emph{represented} by
      an image current $I\,\yhat$ at $x/a=0; z/a=-1$, while $\Bcl$ is
      \emph{represented} by a physical current $I\,\yhat$ at $x/a=0; z/a=1$ and an
     image current $-I\,\yhat$ at $x/a=0; z/a=-1$.
      \label{fig:Decomp}}
\end{figure}  
The gauge non-invariance of the magnetic helicity $H$ is closely related to
the flux threading the bounding surface $\Surf$. This flux is often
mis-attributed to `exterior linkage' similar to the way the potential field
$\BPot$ is often confused with `external linkage' \cite[see pp 30-31
  in][]{Blackman2014}.  Consider the Cartesian
$\left(\xhat,\yhat,\zhat\right)$ geometry of Figure~\ref{fig:Decomp}
where $\Vol=\left(x,y,z>0\right)$ corresponds to the domain of interest
bounded by $\Surf$ at $z=0$. Let this domain $\Vol$ contain a line current
$I\,\yhat$ at $x/a=0$ and $z/a=1$. Figure~\ref{fig:B} shows the physical
current source indicated by the red dot and contours of the vector potential
tracing the magnetic field lines. The color scale represents $B_z$ at $z=0$
(the normal component at $\Surf$) with red being positive and blue
negative. \emph{All of the physical current sources in this example are
contained inside the volume of interest $\Vol$!} Using the generalized
``method of images'' \cite[]{Thomson1845,Hammond1960} any magnetic field $\B$
in $\Vol$ may be decomposed into two components $\B=\BPot+\Bcl$: a magnetic
field $\BPot$ that is potential in $\Vol$ and threads the boundary and a
magnetic field $\Bcl$ that closes on itself in $\Vol$, i.e.,
$\nhat\cdot\Bcl|_\Surf=0$. For example, one can always find a mathematically
unique potential field $\BPot$ in $\Vol$ corresponding to the normal component
of $\B$ on the surface $z=0$, i.e., corresponding to the flux threading this
bounding surface. This potential field, shown in Figure~\ref{fig:BP}, is
represented by an image current $I\,\yhat$ \emph{outside} the volume of
interest $\Vol$ at $x/a=0$ and $z/a=-1$ in $\VolC$, as $\grad\cross\BPot=0$
must be zero inside $\Vol$. Thus, the representation of flux threading the
boundary by a potential field $\BPot$ misattributes the origin of this flux to
a current source outside the volume of interest $\Vol$ \cite[]{Schuck2022}.
Similarly, the magnetic component that closes in $\Vol$, defined as
$\Bcl\equiv\B-\BPot$ and shown in Figure~\ref{fig:BC}, is represented by two
anti-parallel currents: one corresponds to the physical current $I\,\yhat$ in
$\Vol$ and the other corresponds to its image current $-I\,\yhat$. The image current is
symmetrically placed across the boundary $\Surf$ to ensure
$\nhat\cdot\Bcl|_\Surf=0$ at $z=0$ by construction. While mathematically
$\BPot$ is an `external linkage,' its \emph{physical origin} is
\emph{internal} to $\Vol$!  The superposition of $\BPot$ and $\Bcl$ recovers
the total magnetic field because the image currents in $\VolC$ cancel
and all that remains is the physical current source inside $\Vol$.  However,
the decomposition for this example results in the \emph{apparent non-sequitur}
that the potential field $\BPot$ is curl free $\grad\cross\BPot=0$ in $\Vol$
but indeed physically produced by currents in $\Vol$! This example shows how
easily the origins of magnetic fields can be confused by expressing them in
forms that are mathematically convenient, for example, for calculating
relative helicity. Yet the origins of these fields are of critical importance
for understanding cause and effect, and so a means for tracking these origins
while simultaneously calculating global quantities such as the relative
helicity or magnetic energy is key for a complete understanding of dynamical
astrophysical phenomena.\par
The primary purpose of this paper is to disentangle the linkages that
originate with internal and external current sources in relative
helicity for open systems.  This work is organized as follows:
\S\ref{sec:Helmholtz} establishes the framework for attributing
magnetic fields to electric current sources, \S\ref{sec:Helicity}
briefly discusses helicity in \emph{magnetically closed systems},
\S\ref{sec:Relative} reviews the concepts of relative helicity for
\emph{magnetically open systems}, \S\ref{sec:Attribution} extends
relative helicity to simultaneously characterize the
open-open and open-closed helicities as well as
the domains of origin of the linked magnetic fields and
  develops novel expressions for internal and external relative helicity
  and internal-external relative helicity based on the domain of origin
  of the magnetic field in currents, \S\ref{sec:Energy} describes
some of the implications of this work for the concept of free energy
and \S\ref{sec:Conclusions} discusses the implications of these
results for theory and observation.

\section{The Attribution of Magnetic of Field to Internal and External Current Sources \label{sec:Helmholtz}}
The attribution of magnetic fields to physical current sources is
necessary to fully understand cause and effect, the linkages of
helicity, and changes in magnetic energy within a volume of interest
$\Vol$. In classical electromagnetic theory, currents create magnetic
fields. This statement is inherent in the Biot-Savart law
\cite[][]{Biot1820} written in continuous form
\begin{equation}
\B\left(t,\x\right)=\frac{1}{c}\,\grad\cross\int\limits_{\Rthree}{d^3x'}\,\frac{\JM\left(t,\x'\right)}{\left|\x-\x'\right|}=\frac{1}{c}\,\int\limits_{\Rthree}{d^3x'}\,\JM\left(t,\x'\right)\cross\frac{\x-\x'}{\left|\x-\x'\right|^{3}}\qquad\x\in\Rthree,\label{eqn:BS}
\end{equation}
as a convolution with spatial moments of the free space Green's function,
where $c$ is the speed of light.  In the right-most expression there are no
spatial or temporal derivatives operating on the source $\J$.  Green's
functions form the basis for understanding cause and effect in
physics. Loosely speaking, the Green's function propagates a ``cause'' at
$\x'$ to an ``effect'' at $\x$.  This is how nature works despite the practice
in \MHD{} analysis to substitute
$\J\Longrightarrow{c}\,\grad\cross\B/\left(4\,\pi\right)$ into the
$\J\cross\B$ force to eliminate any explicit reference to $\J$ in
\MHD{}. Physically, the current $\J$ is manifestly the \emph{source} of the
magnetic vorticity.  The Biot-Savart law provides attribution of a current
element at $\x'$ to the magnetic field at the location $\x$.  In the
pre-Maxwell formulation of electrodynamics, the magnetic field $\B$ at $\x$
depends on currents at all other points in the universe
$\Rthree$. Realistically, this universe dynamically corresponds to
$\Real\ll{c}\,\Delta t$. This has deep implications\----the magnetic field is
a non-local field despite the fact that it is often conceptually treated as a
local object in \MHD{}. Changes in $\B\left(t,\x\right)$ imply changes in
$\J\left(t,\x'\right)$ somewhere else!  \par
While Equation~(\ref{eqn:BS}) is intuitive, it is nearly
impossible to apply in practice because access to complete
information about all currents in the entire universe $\x\in\Rthree$
is rare. Rather, in most cases, knowledge is limited to currents and
magnetic fields in a volume $\Vol$ bounded by a surface $\Surf$.
Consider a simply connected \emph{internal} volume $\Vol$ bounded by
closed surface $\Surf$ and an \emph{external} domain denoted $\VolC$
such that $\Rthree=\Vol+\VolC$.  Suppose that both domains contain
corresponding current systems $\J$ and $\JME$. By the electromagnetic
superposition principle, the total magnetic field in
Equation~(\ref{eqn:BS}) is then
\begin{subequations}
\begin{equation}
  \B\left(t,\x\right)=\bBS\left(\J;t,\x\right)+\bBS\left(\JME;t,\x\right)\qquad\x\in\Rthree,\label{eqn:B:All}
\end{equation}  
with
\begin{align}  
\bBS\left(\J;t,\x\right)=\overbrace{\frac{1}{c}\,\grad\cross\int\limits_{\Vol}{d^3x'}\,\frac{\JM\left(t,\x'\right)}{\left|\x-\x'\right|}}^{\mbox{Internal
      Sources}}\qquad\x\in\Rthree,\label{eqn:BS:JM}\\
\bBS\left(\JME;t,\x\right)=\overbrace{\frac{1}{c}\,\grad\cross\int\limits_{\VolC}{d^3x'}\,\frac{\JME\left(t,\x'\right)}{\left|\x-\x'\right|}}^{\mbox{External
      Sources}}\qquad\x\in\Rthree,\label{eqn:BS:JME}
\end{align}  
\end{subequations}
where the total field is comprised of two \emph{integrants}: one
produced by \emph{internal} sources, $\JM$ in $\Vol$, and one produced
by \emph{external} sources, $\JME$ in $\VolC$. Both
integrants~(\ref{eqn:BS:JM}) and (\ref{eqn:BS:JME}) are continuous
vector fields for $\x\in\Rthree$. If $\JM$ and $\B$ are completely
known in $\Vol$ then $\bBS\left(\JM;t,\x\right)$ can be computed
directly by convolution and $\bBS\left(\JME;t,\x\right)$ in $\Vol$ may
be computed from Equation~(\ref{eqn:B:All}). Analogously, if $\B$ in
$\Vol$ is known and $\bBS\left(\JME;t,\x\right)$ can be estimated then
$\bBS\left(\JM;t,\x\right)$ in $\Vol$ may be computed from
Equation~(\ref{eqn:B:All}). Below we show that
$\bBS\left(\JM;t,\x\right)$ and $\bBS\left(\JME;t,\x\right)$ in $\Vol$
and $\bBS\left(\JM;t,\x\right)$ in $\VolC$ may be computed from $\B$
in $\Vol\cup\Surf$ \emph{without} performing computationally expensive
Biot-Savart convolution integrals by leveraging the powerful
fundamental theorem of vector calculus. For the remainder of the paper
we include the source $\JM$ or $\JME$ as an argument to the vector
field when the source of the magnetic field is of interest. For
example, $\BPot\left(\JM;t,\x\right)$ and $\Bcl\left(\JM;t,\x\right)$
are, respectively, the potential magnetic field and the magnetic field
that closes in $\Vol$ determined from the magnetic field
$\bBS\left(\JM;t,\x\right)$ produced by currents $\JM$ in $\Vol$.
Correspondingly $\BPot\left(\JME;t,\x\right)$ and
$\Bcl\left(\JME;t,\x\right)$ are, respectively, the potential magnetic
field and the magnetic field that closes in $\Vol$ determined from the
magnetic field $\bBS\left(\JME;t,\x\right)$ produced by currents
$\JME$ in $\VolC$.  And finally, $\B\left(t,\x\right)$ without the
argument of current represents the total magnetic field at $t$ and
$\x$ in Equation~(\ref{eqn:B:All}).  \par
\subsection{The Fundamental Theorem of Vector Calculus: The Helmholtz Decomposition}
Consider the fundamental theorem of vector calculus (the Helmholtz
Decomposition, \HD) for a vector field $\Field\left(\x\right)$ in $\Vol$ \cite[]{Morse1953a,Gui2007,Kustepeli2016}
\begin{subequations}
\begin{equation}  
\FieldHD\left(\x\right)=\grad\cross\AV\left(\x\right)-\grad\Psi\left(\x\right),\label{eqn:HT}
\end{equation}  
where
\begin{align}
\AV\left(\x\right)=&\frac{1}{4\,\pi}\,\left[\int\limits_\Vol{\!d^3x'}\frac{\grad'\cross\Field\left(\x'\right)}{\left|\x-\x'\right|}+\oint\limits_\Surf{dS'}\,\frac{\nhat'\cross\Field\left(\x'\right)}{\left|\x-\x'\right|}\right],\\
\Psi\left(\x\right)=&\frac{1}{4\,\pi}\,\left[\int\limits_\Vol{\!d^3x'}\frac{\grad'\cdot\Field\left(\x'\right)}{\left|\x-\x'\right|}+\oint\limits_\Surf{dS'}\,\frac{\nhat'\cdot\Field\left(\x'\right)}{\left|\x-\x'\right|}\right],\label{eqn:Psi}
\end{align}  
where $\nhat$ points into $\Vol$. There is a jump discontinuity in the
value of the surface integrals as the \emph{observation point} $\x$
passes from $\Vol$ to $\VolC$ across a smooth surface $\Surf$ producing
\begin{equation}
  \FieldHD\left(\x\right)=\left\lbrace\begin{array}{ll}
  \Field\left(\x\right)\qquad&\x\in\Vol,\\
  \Field\left(\x\right)/2\qquad&\x\in\Surf,\\
  0&\x\in\VolC.
  \end{array}\right.\label{eqn:FieldHD}  
\end{equation}  
\end{subequations}
The \HD{} is a mathematical reconstruction theorem. It is ignorant of
electromagnetic theory and does not inherently preserve physical
properties of the field $\Field\left(\x\right)$ across the boundary
$\Surf$. For example, if $\Field$ is solenoidal for $\x\in\Rthree$,
then generally Equation~(\ref{eqn:FieldHD}) will not maintain this
property, e.g., continuity of $\nhat\cdot\Field$, across
$\Surf$. Furthermore, its value on a smooth surface converges to half
the value just inside the boundary, which is an inconvenient property
for astrophysical problems that involve physics in notional surfaces
between domains, such as a photosphere.  This motivates the
alternative definition \cite[see for example][]{Kempka1996}
\begin{subequations}
\begin{equation}  
\Factor\left(\x\right)\FieldGHD\left(\x\right)=\grad\cross\AV\left(\x\right)-\grad\Psi\left(\x\right),\label{eqn:HTK}
\end{equation}  
\begin{equation}
  \Factor\left(\x\right)=\frac{\chi\left(\x\right)}{4\,\pi}=\left\lbrace\begin{array}{ll}
  1\qquad&\x\in\Vol\\
 \left. \begin{array}{ll}1/2&\mbox{smooth surfaces}\\
    1/4&\mbox{edges of $\Vol$}\\
  1/8&\mbox{vertices of $\Vol$}\end{array}\right\rbrace\qquad&\x\in\Surf\\
  0\qquad&\x\in\VolC
\end{array}\right.,\label{eqn:Alpha}  
\end{equation}
where $\chi\left(\x\right)$ is the local internal solid angle of the
\emph{principal volume} at the observation point on $\Surf$
\cite[]{Kellogg1929,Courant1989a,Courant1989b,Bladel1991}. The factor
$\Factor\left(\x\right)$ is a constant, and therefore continuous and
differentiable, on the open sets $\x\in\Vol$ and $\x\in\VolC$ which do not
contain $\x\in\Surf$. The factor $\Factor\left(\x\right)$ takes on other
values when $\x$ lies in the boundary $\Surf$ because the principle volume
of the observation point projects into both domains $\x\in\Vol$ and
$\x\in\VolC$.  On smooth boundaries $\Surf$ with well-defined tangent
surfaces $\Factor=1/2$, i.e., half the principle volume lies in $\Vol$ and
half in $\VolC$. By analogy, for a cube, which is smooth almost everywhere,
$\Factor=1/2$ on faces, $\Factor=1/4$ on edges, and $\Factor=1/8$ at vertices
(of a cuboid) and of course $\Factor=1$ for $\x\in\Vol$ and $\Factor=0$ for
$\x\in\VolC$, consistent with the projections of the fractions of the
principal volumes into $\Vol$.\par
The $\Factor\left(\x\right)$ on the left of Equation~(\ref{eqn:HTK})
ensures that the surface values of $\FieldGHD\left(\x\right)$ are
continuous from within the volume $\Vol$ as defined by the
one-sided limiting process
\begin{equation}
\lim_{\x\in\Vol\rightarrow\x\in\Surf}\FieldGHD\left(\x\right)=\Field\left(\x\right).\label{eqn:Continuity}  
\end{equation}  
Consequently
\begin{equation}
  \FieldGHD\left(\x\right)=\left\lbrace\begin{array}{ll}
  \Field\left(\x\right)\qquad&\x\in\Vol\cup\Surf,\\
  \mbox{Arbitrary}&\x\in\VolC.
  \end{array}\right.\label{eqn:FieldGHD}  
\end{equation}  
\end{subequations}
$\FieldGHD\left(\x\right)$ is arbitrary in $\VolC$ because
$\Factor\left(\x\right)=0$ on the left-hand side of
Equation~(\ref{eqn:HTK}) for $\x\in\VolC$. Thus,
$\FieldGHD\left(\x\right)$ can formally be defined in $\VolC$ to
properly preserve physical properties of $\Field\left(\x\right)$
across $\Surf$.\par
For a solenoidal field, the divergence term in~(\ref{eqn:Psi}) may be ignored
and expression $\FieldGHD\left(\x\right)$ for the magnetic field becomes
\begin{equation}
\Factor\left(\x\right)\,\BHD\left(t,\x\right)=
\frac{1}{4\,\pi}\,\grad\cross\int\limits_\Vol{d^3x'}\frac{\grad'\cross\B\left(t,\x'\right)}{\left|\x-\x'\right|}+\frac{1}{4\,\pi}\,\left[\grad\cross\oint\limits_\Surf{dS'}\,\frac{\nhat'\cross\B\left(t,\x'\right)}{\left|\x-\x'\right|}-\grad\oint\limits_\Surf{dS'}\,\frac{\nhat'\cdot\B\left(t,\x'\right)}{\left|\x-\x'\right|}\right]\qquad\x\in\Rthree.\label{eqn:B:V}
\end{equation}  
As mentioned above, this does not constrain $\BHD$ in the external universe
where $\Factor\left(\x\right)=0$.  Strictly speaking, if there is flux
threading the boundary $\Surf$ then the magnetic field determined by
Equation~(\ref{eqn:B:V}) for $\x\in\Vol\cup\Surf$ should be formally
matched to a potential field in the external universe $\x\in\VolC$ to
preserve the solenoidal property of $\B$ across $\Surf$, i.e., as discussed in
relation to Equation~(\ref{eqn:FieldGHD}) above for $\x\in\VolC$. However,
practically speaking, this matching procedure is usually unnecessary as we are
often interested in reconstructing \begin{enumerate*}
\item $\B$ for $\x\in\Vol\cup\Surf$ \label{itm:Recon} or
  determining \item $\bBS\left(\JME;t,\x\right)$ for
  $\x\in\Vol\cup\Surf$ \label{itm:External} or \item
  $\bBS\left(\JM;t,\x\right)$ for $\x\in\Rthree$  \label{itm:Internal} as discussed
  below. \end{enumerate*} \par
\subsection{Linking Magnetic Fields to their Current Sources}
If the \emph{net} displacement current is ignorable, then Amp{\`e}re's
law
\begin{equation}
\grad\cross\B=\frac{4\,\pi}{c}\,\J,\label{eqn:Ampere}
\end{equation}  
may be substituted into the volume integral to produce 
\begin{equation}
\Factor\left(\x\right)\,\BHD\left(t,\x\right)=
\frac{1}{c}\,\grad\cross\int\limits_\Vol{d^3x'}\frac{\J\left(t,\x'\right)}{\left|\x-\x'\right|}+\frac{1}{4\,\pi}\,\left[\grad\cross\oint\limits_\Surf{dS'}\,\frac{\nhat'\cross\B\left(t,\x'\right)}{\left|\x-\x'\right|}-\grad\oint\limits_\Surf{dS'}\,\frac{\nhat'\cdot\B\left(t,\x'\right)}{\left|\x-\x'\right|}\right]
\qquad\x\in\Rthree.\label{eqn:GHD:J}
\end{equation}
Equation~(\ref{eqn:B:All}) unambiguously associates the Biot-Savart integrals
over current systems $\JM$ and $\JME$ to their corresponding magnetic field
components $\bBS\left(\JM;t,\x\right)$ and $\bBS\left(\JME;t,\x\right)$,
establishing cause and effect.  This pre-Maxwell equation also implies that
the magnetic field $\B\left(t,\x\right)$ at any location contains
entangled magnetic contributions from both internal $\JM$ and external
$\JME$ current systems. Thus, the surface integrals in
Equation~(\ref{eqn:GHD:J}) \emph{implicitly} also contain entangled
magnetic contributions from both internal $\JM$ and external $\JME$ current
systems. As shown below, these contributions separate cleanly when
$\x\in\Vol$ or $\x\in\VolC$ but are entangled when the observation point is in
the boundary $\x\in\Surf$. \par
Since the factor $\Factor\left(\x\right)$ is chosen to enforce
continuity of the \HD{} from $\Vol$ to $\Surf$, as in
Equation~(\ref{eqn:Continuity}), the discussion of (\ref{eqn:GHD:J}) is
divided logically into two domains $\x\in\Vol\cup\Surf$ and
$\x\in\VolC$. For $\x\in\Vol\cup\Surf$, Equation~(\ref{eqn:GHD:J}) becomes
\begin{equation}
\B\left(t,\x\right)=
\frac{1}{\Factor\left(\x\right)\,c}\,\grad\cross\int\limits_\Vol{d^3x'}\frac{\J\left(t,\x'\right)}{\left|\x-\x'\right|}+\frac{1}{4\,\pi\,\Factor\left(\x\right)}\,\left[\grad\cross\oint\limits_\Surf{dS'}\,\frac{\nhat'\cross\B\left(t,\x'\right)}{\left|\x-\x'\right|}-\grad\oint\limits_\Surf{dS'}\,\frac{\nhat'\cdot\B\left(t,\x'\right)}{\left|\x-\x'\right|}\right]
\qquad\x\in\Vol\cup\Surf,\label{eqn:GHD:B}
\end{equation}
which addresses the reconstruction in item~\ref{itm:Recon}
above. Equations~(\ref{eqn:B:All}) and~(\ref{eqn:GHD:B}) are
equivalent in the intersection of their domain of validity
$x\in\Vol\cup\Surf$. This equivalence will be used to establish the
formal correspondence between $\bBS\left(\JM;t,\x\right)$ and
$\bBS\left(\JME;t,\x\right)$ in
Equations~(\ref{eqn:BS:JM})\--(\ref{eqn:BS:JME}) and the \HD{} in
Equation~(\ref{eqn:GHD:B}).  \par
To establish the correspondence between internal and external sources
in Equation~(\ref{eqn:B:All}) and terms in~(\ref{eqn:GHD:B}), the Biot-Savart
magnetic field produced by \emph{internal sources} in the volume
$\Vol$ from Equation~(\ref{eqn:BS:JM}) is added to and subtracted from
Equation~(\ref{eqn:GHD:B}) to produce for $\x\in\Vol\cup\Surf$
\begin{align}
\B\left(t,\x\right)=&
\underbrace{\overbrace{\frac{1}{c}\,\grad\cross\int\limits_\Vol{\!d^3x'}\frac{\J\left(t,\x'\right)}{\left|\x-\x'\right|}}^{\mbox{Internal
    Sources}}}_{\bBS\left(\JM;t,\x\right)}\label{eqn:BGHD:ALL}\\
&\qquad+\underbrace{\overbrace{\frac{1-\Factor\left(\x\right)}{\Factor\left(\x\right)\,c}\,\grad\cross\int\limits_\Vol{\!d^3x'}\frac{\J\left(t,\x'\right)}{\left|\x-\x'\right|}+\frac{1}{4\,\pi\,\Factor\left(\x\right)}\,\left[\grad\cross\oint\limits_\Surf{dS'}\,\frac{\nhat'\cross\B\left(t,\x'\right)}{\left|\x-\x'\right|}-\grad\oint\limits_\Surf{dS'}\,\frac{\nhat'\cdot\B\left(t,\x'\right)}{\left|\x-\x'\right|}\right]}^{\mbox{External  Sources}}}_{\bBS\left(\JME;t,\x\right)}, \nonumber 
\end{align}  
where the terms are now grouped according to their physical
interpretation. This resolves items~\ref{itm:External} and \ref{itm:Internal}
for $\x\in\Vol\cup\Surf$. However, as discussed below, there are more
efficient computational expressions for $\bBS\left(\JM;t,\x\right)$ and
$\bBS\left(\JME;t,\x\right)$ when the bounding surface is excluded, i.e.,
$\x\in\Vol$ or $\x\in\VolC$.  Note that the formal appearance of the integrant
due to internal sources proportional to $\bBS\left(\JM;t,\x\right)$ under
``External Sources'' in Equation~(\ref{eqn:BGHD:ALL}) is a consequence of the
entanglement of internal and external sources of $\B$ in evaluation of
the surface integrals \emph{when the observation point is in the boundary
$\x\in\Surf$.} The surface integrals depend on total magnetic field
$\B\left(t,x\right)$ which implicitly contains entangled magnetic
contributions from internal and external current sources. If we exclude the
observation points in the surface, then $\Factor\left(\x\right)=1$ and, for
observation points in the volume of interest, Equation~(\ref{eqn:BGHD:ALL})
reduces to the intuitive form
\begin{equation}
\B\left(t,\x\right)=
\underbrace{\overbrace{\frac{1}{c}\,\grad\cross\int\limits_\Vol{\!d^3x'}\frac{\J\left(t,\x'\right)}{\left|\x-\x'\right|}}^{\mbox{Internal
  Sources}}}_{\bBS\left(\JM;t,\x\right)}+\underbrace{\overbrace{\frac{1}{4\,\pi}\,\left[\grad\cross\oint\limits_\Surf{dS'}\,\frac{\nhat'\cross\B\left(t,\x'\right)}{\left|\x-\x'\right|}-\grad\oint\limits_\Surf{dS'}\,\frac{\nhat'\cdot\B\left(t,\x'\right)}{\left|\x-\x'\right|}\right]}^{\mbox{External Sources}}}_{\bBS\left(\JME;t,\x\right)}\qquad\x\in\Vol.\label{eqn:BGHD:V}  
\end{equation}  
The surface integrals now provide an efficient expression for
the magnetic field in $\Vol$ produced by external sources
\begin{subequations}
\begin{align}  
  \bBS\left(\JME;t,\x\right)=&\B\left(t,\x\right)- \bBS\left(\JM;t,\x\right),\label{eqn:B:Surf:Externala}\\
=&\frac{1}{4\,\pi}\,\left[\grad\cross\oint\limits_\Surf{dS'}\,\frac{\nhat'\cross\B\left(t,\x'\right)}{\left|\x-\x'\right|}-\grad\oint\limits_\Surf{dS'}\,\frac{\nhat'\cdot\B\left(t,\x'\right)}{\left|\x-\x'\right|}\right]=\overbrace{\frac{1}{c}\,\grad\cross\int\limits_{\VolC}{d^3x'}\frac{\JME\left(t,\x'\right)}{\left|\x-\x'\right|}}^{\mbox{External
    Sources}}\qquad\x\in\Vol.\label{eqn:B:Surf:External}
\end{align}  
This establishes $\bBS\left(\JME;t,\x\right)$ for $\x\in\Vol$ by
surface convolution alone. This expression may be subtracted from the
total field $\B$ to provide an expression for the internal sources by
surface convolution that is equivalent to the Biot-Savart law for
internal sources
\begin{equation}  
\bBS\left(\JM;t,\x\right)=\B\left(t,\x\right)-\bBS\left(\JME;t,\x\right)=\overbrace{\frac{1}{c}\,\grad\cross\int\limits_{\Vol}{d^3x'}\frac{\JM\left(t,\x'\right)}{\left|\x-\x'\right|}}^{\mbox{Internal
      Sources}}\qquad\x\in\Vol.\label{eqn:B:Surf:External:Internal}
\end{equation}  
\end{subequations}
Analogously, if we consider observation points in the external universe then
$\Factor\left(\x\right)=0$ and the surface terms \emph{extinguish} the
internal terms as Equation~(\ref{eqn:GHD:J}) becomes
\begin{subequations}
\begin{equation}
0=\underbrace{\overbrace{\frac{1}{c}\grad\cross\int\limits_\Vol{d^3x'}\frac{\JM\left(t,\x'\right)}{\left|\x-\x'\right|}}^{\mbox{Internal
  Sources}}}_{\bBS\left(\JM;t,\x\right)}+\underbrace{\overbrace{\frac{1}{4\,\pi}\,\left[\grad\cross\oint\limits_\Surf{dS'}\,\frac{\nhat'\cross\B\left(t,\x'\right)}{\left|\x-\x'\right|}-\grad\oint\limits_\Surf{dS'}\,\frac{\nhat'\cdot\B\left(t,\x'\right)}{\left|\x-\x'\right|}\right]}^{-\left(\mbox{Internal
  Sources}\right)}}_{-\bBS\left(\JM;t,\x\right)}\,\qquad\x\in\VolC,\label{eqn:BGHD:Extinction}  
\end{equation}  
where
\begin{equation}  
\bBS\left(\JM;t,\x\right)=-\frac{1}{4\,\pi}\,\left[\grad\cross\oint\limits_\Surf{dS'}\,\frac{\nhat'\cross\B\left(t,\x'\right)}{\left|\x-\x'\right|}-\grad\oint\limits_\Surf{dS'}\,\frac{\nhat'\cdot\B\left(t,\x'\right)}{\left|\x-\x'\right|}\right]=\overbrace{\frac{1}{c}\,\grad\cross\int\limits_{\Vol}{d^3x'}\frac{\JM\left(t,\x'\right)}{\left|\x-\x'\right|}}^{\mbox{Internal
    Sources}}\qquad\x\in\VolC.\label{eqn:B:Surf:Internal}
\end{equation}  
\end{subequations}
Equations~(\ref{eqn:B:Surf:External}) and~(\ref{eqn:B:Surf:Internal})
manifestly show that the surface integrals contain contributions to the
magnetic field from both internal and external currents and that these
contributions separate out cleanly for observation points $\x\in\Vol$ or
$\x\in\VolC$ but are entangled for $\x\in\Surf$.
Equations~(\ref{eqn:B:Surf:External})\--(\ref{eqn:B:Surf:External:Internal})
and~(\ref{eqn:B:Surf:Internal}) establish $\bBS\left(\JM;t,\x\right)$ for
$\x\in\Vol$ in item~\ref{itm:Internal} by surface convolution alone.  Note
that even if $\bBS\left(\JM;t,\x\right)$ or $\bBS\left(\JME;t,\x\right)$ are
required on $\Surf$, this computation necessitates the evaluation of the
Biot-Savart convolution only for surface points $\x\in\Surf$. Furthermore,
there are other more direct techniques for separating magnetic fields into
$\bBS\left(\JM;t,\x\right)$ or $\bBS\left(\JME;t,\x\right)$ on a closed smooth
surface \cite[see][for a technique applicable to a spherical
  boundary]{Schuck2022}. Recently, \cite{Leake2023} have developed a tool for
applying the \HD{} in Equations~(\ref{eqn:HTK})\--(\ref{eqn:Alpha})
and~(\ref{eqn:B:V}) for astrophysical \MHD{} simulations.\par
Having established this framework for the attribution of magnetic fields to
their origin in internal and external current sources, we now turn our
attention to the implications of this causality for magnetic helicity and
magnetic energy.

\section{Helicity for Magnetically `Closed' Systems \label{sec:Helicity}}
A magnetically `closed' system has no magnetic flux threading the boundary
$\Surf$ anywhere, i.e., $\left.\B\cdot\nhat\right|_\Surf=0$. As
demonstrated below, even a magnetically closed system with
$\nhat\cdot\B|_\Surf=0$ is not completely electrodynamically isolated from the
external universe $\VolC$.\par
For an ideal plasma, the evolution of the vector
potential in the incomplete Gibbs gauge\footnote{This gauge condition
is referenced as the ``Gibbs,'' ``Weyl,'' ``Hamiltonian,'' and
``temporal'' gauge in the literature
\cite[]{Gibbs1896,Przeszowski1996,Jackson2002}.} is determined by
\begin{equation}
\frac{\partial \AV}{\partial t}=\vel\cross\B,\label{eqn:Weyl}  
\end{equation}
where $\vel$ is the plasma velocity. \cite{Woltjer1958} showed
that the magnetic helicity is invariant 
\begin{equation}
\frac{d H}{d t}=\frac{\partial}{\partial t}\int\limits_\Vol{d^3x}\,\AV\cdot\grad\cross\AV=\oint\limits_\Surf{dS}\,\nhat\cdot\AV\cross\frac{\partial\AV}{\partial t}=0,\label{eqn:Woltjer}
\end{equation}
in a closed system, stating
\begin{quote}
  The surface integral vanishes because we consider a closed
  system. For then the motions inside the system may not affect the
  vector potential outside, and, as the vector potential is
  continuous, even when surface currents are present,
  $\partial\AV/\partial t$ must vanish at the surface of the system.
\end{quote}
\begin{figure}
  \centering \setlength{\mywidth}{1.25in} \centering \subfloat[A magnetically closed system:
    $z>0$]{\includegraphics[width=\mywidth,angle=90]{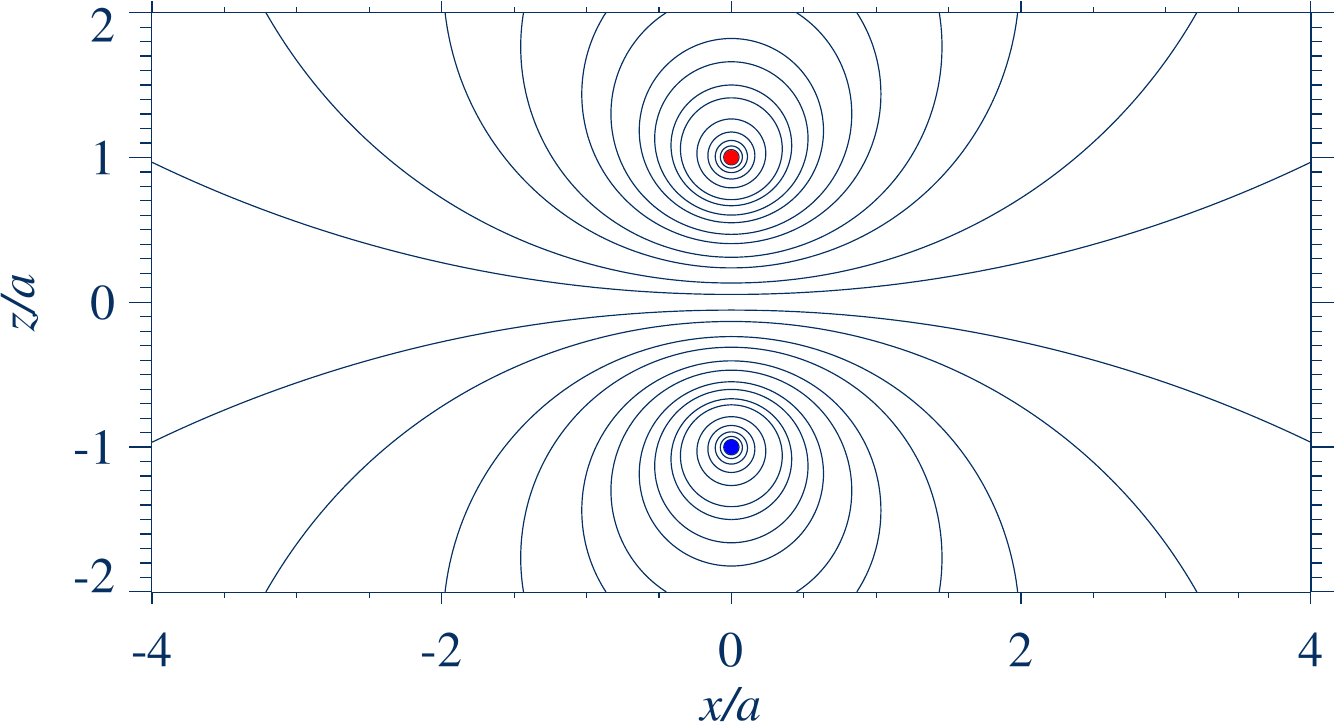}
      \label{fig:BP12}}
    \subfloat[Internal current system: $z>0$]{
      \includegraphics[width=\mywidth,angle=90,viewport=0 0 354 580
        ,clip=]{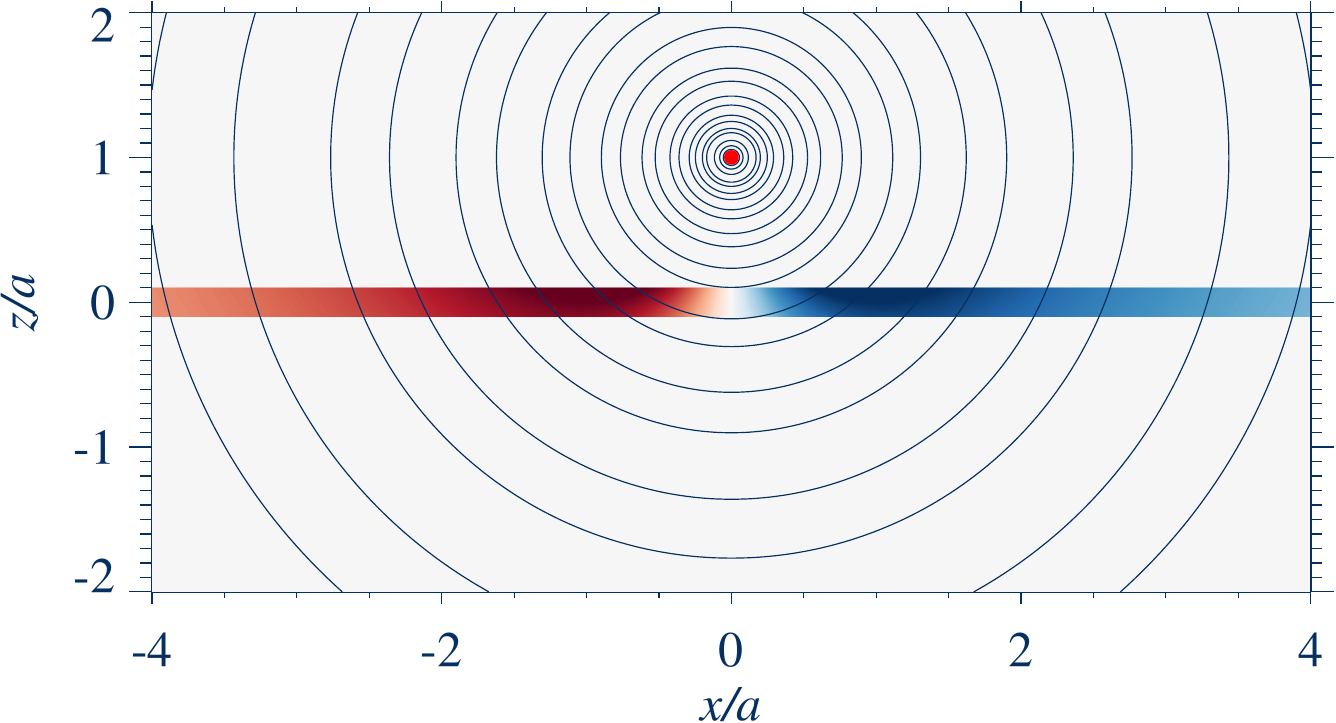}
    \label{fig:BP:Internal}}
    \subfloat[External current system: $z<0$]{
    \includegraphics[width=\mywidth,angle=90,viewport=0 0 354 580 ,clip=]{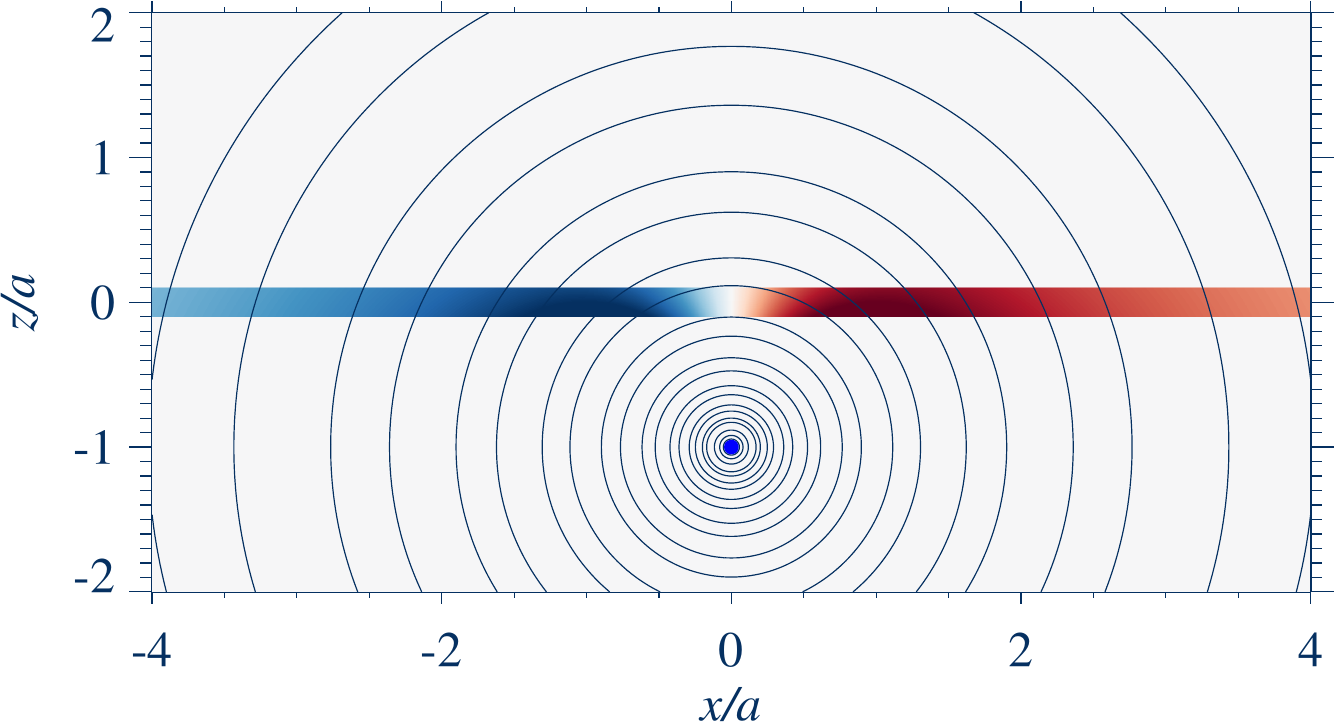}
    \label{fig:BP:External}}
    \caption{The entangled origins of \subref{fig:BP12} the intrinsically
      solenoidal magnetic field $\Bcl$ of a magnetically closed system
      for $z>0$ with $\nhat\cdot\Bcl|_\Surf=0$ that is produced by two
      physical current systems: \subref{fig:BP:Internal} one internal,
      $I\left(t\right)\,\yhat$ at $x/a=0$ and $z/a=1$, and
      \subref{fig:BP:External} one external, $-I\left(t\right)\,\yhat$ at
      $x/a=0$ and $z/a=-1$.  A red dot indicates a line current directed away
      from the observer and a blue dot indicates a line current towards the
      observer. The black lines are contours of the vector potential that
      trace magnetic field lines. The color scale along $z=0$ corresponds to
      the vertical magnetic field component with red/blue corresponding to
      up/down.
      \label{fig:Closed}}
\end{figure}  
The physical
implications of these boundary conditions are discussed in more detail in
Appendix~\ref{sec:Boundary}.  However, we touch on some obvious points
here. First, $\partial\AV/\partial t|_\Surf=0$ implies that
$\nhat\cdot\partial\B/\partial t|_\Surf=0$ but does not imply either
  $\nhat\cdot\J|_\Surf=0$ or $\nhat\cdot\partial\J/\partial
  t|_\Surf=0$\----the two domains may share time-dependent current
  systems that pass through $\Surf$.\footnote{Consider the Cartesian
  example with $\Surf$ defined as $z=0$ with $\nhat=\zhat$
  \begin{displaymath}
{\partial_t \AV}=\grad\cross\left(\partial_t\psi\,\zhat\right)+\partial_t{A}_z\,\zhat+\grad_\surf\partial_t\phi\qquad\mbox{with}\left.\partial_t \AV\right|_\Surf=0\iff\partial_t\psi|_\Surf=\mbox{constant, }\partial_t{A}_z|_\Surf=0\mbox{ and }\partial_t\phi|_\Surf=\mbox{constant}
    \end{displaymath}
and $\grad_\surf\equiv\xhat\,\partial_x+\yhat\,\partial_y$ and where we have
used the short-hand $\partial_x=\partial/\partial x$ in this footnote.  Then
    \begin{displaymath}
    \left.\zhat\cdot\grad\cross\grad\cross\partial_t \AV\right|_\Surf=\left.\nabla^2_\surf\partial_z\partial_t\phi\right|_\Surf\neq0,   
    \end{displaymath}
which is not required to be zero. Note that strictly speaking the vector
potential must also satisfy $\B\cdot\partial_t\AV=0$, i.e., Equation~(\ref{eqn:Weyl}). For example, consider the case where there is a cylindrically symmetric vertical current $I\,\zhat$ passing through the domain, generating an azimuthal $B_\theta$ in the domain, with no other magnetic field. This field has no linkages, and so no helicity. Even if the current amplitude is changed, the system remains in a zero helicity state.} Second, hidden in Equation~(\ref{eqn:Woltjer})
is the assumption of gauge invariance which requires $\nhat\cdot\B|_\Surf=0$,
e.g., $\nhat\cross\AV|_\Surf=0$ to within a gauge transformation (see
Appendix~\ref{sec:Boundary}). This is a stronger assumption than
$\nhat\cdot\partial\B/\partial t|_\Surf=0$. Third, within the volume of
interest $\Vol$, the magnetic field produced by ``surface currents'' is
electrodynamically indistinguishable from the magnetic field produced by
external currents $\JME$ in $\VolC$. This last point suggests that even with
the mathematical boundary conditions imposed by \cite{Woltjer1958} that under
special, albeit contrived, conditions, the motions in $\VolC$ can affect the
magnetic field and plasma in $\Vol$.  For example, consider the pedagogical
system $\Vol$ and $\VolC$ bounded by $\Surf$ at $z=0$ shown in
Figure~\ref{fig:Closed}. The vector potential of this system is given by
\begin{equation}  
A_y\left(t,x,z\right)=-\frac{I\left(t\right)}{c}\,\log\left[\frac{x^2+\left(z-a\right)^2}{x^2+\left(z+a\right)^2}\right],
\end{equation}  
where $I(t)$ is the current in the two thin, oppositely directed current
channels at $x/a=0$ and $z/a=\pm1$. The normal component, $B_z$, which is a
superposition of these two current sources, has been contrived to precisely
cancel at $z=0$, and $A_y(t,x,z=0)=0$ and thus $\Vol$ is a magnetically closed
system by the mathematical boundary condition $\partial\AV/\partial
t|_\Surf=0$ in \cite{Woltjer1958}.  While there is no flux threading the
boundary $\Surf$ at $z=0$, Figure~\ref{fig:BP:Internal} shows that flux
produced by the physical current source $I\left(t\right)\,\yhat$ at $x/a=0$
and $z/a=1$ threads the boundary and permeates $\VolC$.  Similarly
Figure~\ref{fig:BP:External} shows that flux produced by the physical current
source $-I\left(t\right)\,\yhat$ at $x/a=0$ and $z/a=-1$ threads the boundary
and permeates $\Vol$. The total magnetic field $\B\left(t,\x\right)$ for
$\Vol$ shown in Figure~\ref{fig:BP12} entangles the magnetic field from these
two physical current sources and thus $\Vol$ and $\VolC$ are ``communicating''
in collusion to satisfy $B_z=0$ at $z=0$.  This system results in the apparent
non-sequitur that there can be magnetic field in $\Vol$ produced by currents
$\JME$ in $\VolC$ and magnetic field in $\VolC$ produced by currents $\JM$ in
$\Vol$ when no magnetic flux threads $\Surf$, the boundary between $\Vol$ and
$\VolC$.  Of course this highly idealized system is not in force balance and
is likely to relax violently to a lower energy state.  Nonetheless, this
example serves to demonstrate that a magnetically `closed' system is not
necessarily electrodynamically isolated and may be implicitly coupled to
the external universe. \emph{A magnetically closed system may contain
magnetic field produced by the external universe.}  Furthermore, the
collusion between $\Vol$ and $\VolC$ described here is in common use in
solar physics. It is analogous to the collusion between $\Vol$ and
$\VolC$ required to impose flux preserving boundary conditions
($\nhat\cdot\partial \B/\partial t|_\Surf=0$) on the photosphere in
photosphere-to-corona \MHD{} simulations of the solar atmosphere
\cite[e.g.,][]{Kusano1995,Knizhnik2017b,Linan2020,Bian2023}.  We remark that
the remainder of this paper is devoted to understanding the situation where
two systems $\Vol$ and $\VolC$ are magnetically open and manifestly
electrodynamically coupled.\par
\cite{Woltjer1958} further showed that a force-free field $\J=\lambda\,\B$
with constant $\lambda$ represents the lowest state of magnetic energy that a
magnetically closed system containing an ideal plasma can achieve while
constrained by a prescribed helicity $H$. However, there was no obvious
pathway for an ideal plasma to relax to the \citeauthor{Woltjer1958}
state, because the equations of motion for an ideal \MHD{} plasma exhibit an infinite
number of symmetries corresponding to dynamical invariants, by
\citeauthor{Noether1918b}'s (\citeyear{Noether1918b}) first theorem
\cite[]{Frenkel1982}.\footnote{This is sometimes called
\citeauthor{Noether1918b}'s (\citeyear{Noether1918b}) second theorem, but see
footnote~\ref{ft:Noether} and \cite{Brading2000} and \cite{Brading2002}. }
Thus, while the magnetic helicity, $H$, is preserved in an ideal plasma,\footnote{See also \cite{Moffatt1969} and pp. 44-45
  in \cite{Moffatt1978} for a different approach to helicity conservation.} it
is not particularly unique or useful for describing ideal plasma dynamics\----it is invariant, but it is just one of the infinity of invariants.  The situation is different for a non-ideal plasma because
\cite{Taylor1974} conjectured that the magnetic helicity $H$ remains invariant
even in the presence of weak dissipation which destroys the conservation of
the other quantities. The \citeauthor{Taylor1974} conjecture provided a
pathway for a closed system containing a near-ideal plasma to relax to the
\citeauthor{Woltjer1958} linear force-free state while constrained by a
prescribed helicity $H$.  Helicity is a so-called ``robust invariant,''
meaning that it is approximately preserved during a rapid plasma relaxation to
a lower energy state even if that involves dissipation, reconnection, and
magnetic reorganization.  Thus, while challenging to quantify, the magnetic
helicity is an important measure of magnetic complexity in a near-ideal
plasma.  \par

\section{Relative Helicity for Magnetically `Open' Systems\label{sec:Relative}}
The concept of helicity was then extended, by \cite{Berger1984a} and
\cite{Finn1985}, from magnetically `closed' systems where magnetic
``communication'' is limited to current systems $\JM$ in $\Vol$ and $\JME$ in
$\VolC$ that act in collusion to preserve $\nhat\cdot\B|_\Surf=0$ to
magnetically `open' systems where magnetic communication is manifest because
magnetic flux threads the boundary $\Surf$ and $\JM$ in $\Vol$ and $\JME$ in
$\VolC$ can each independently contribute to $\nhat\cdot\B|_\Surf$.  In this
context, the helicity is measured relative to a \emph{reference magnetic
field} $\BRef$ which threads the boundary $\Surf$ of $\Vol$ in the same way as
the magnetic field $\B$.  Irregardless, for either open or closed systems, the
linkages produced by currents in the external universe $\VolC$ can become
entangled with the linkages produced by currents in the internal volume
$\Vol$. The goal of this paper is to extend the work of \cite{Woltjer1958},
\cite{Berger1984a}, \cite{Finn1985}, and \cite{Berger1999,Berger2003} and
provide a clear distinction between the origin of the linkages in $\Vol$.\par
The relative magnetic helicity measure for systems supporting magnetic
fields that thread the boundary $\Surf$ proposed by \cite{Berger1984a}
and \cite{Finn1985} is
\begin{equation}
\Helicity=\int\limits_\Vol{d^3x}\left(\AV+\ARef\right)\cdot\left(\B-\BRef\right),\label{eqn:HR} 
\end{equation}
where the reference magnetic field $\BRef$ that threads the boundary
$\Surf$ is defined as
\begin{equation}
\BRef=\grad\cross\ARef\qquad\mbox{where}\qquad\left.\left(\B-\BRef\right)\cdot\nhat\right|_\Surf=0,
\end{equation}
and the magnetic field that closes in $\Vol$  is then defined as
\begin{subequations}
\begin{align}
\Acl=&\AV-\ARef,\\  
\Bcl=&\B-\BRef=\grad\cross\Acl.\label{eqn:Bcl}
\end{align}  
\end{subequations}
The reference field represents the `open' magnetic field that threads $\Surf$
because $\nhat\cdot\BRef$ is nonzero on $\Surf$, i.e., $\BRef$ has components
that enter and leave $\Vol$. The magnetic field $\Bcl$ is solenoidal
  $\grad\cdot\Bcl=0$, closes on itself in $\Vol$, and thus exhibits no normal
component on $\Surf$\----it is an \emph{intrinsically solenoidal} vector
  field in $\Vol$ \cite[]{Kemmer1977,Schuck2019}. Note that
expression~(\ref{eqn:HR}) is gauge invariant because any gauge transformation
of $\AV$ or $\ARef$ will involve the integral of the dot product between the
gradient of a scalar $\grad\gauge$ and an intrinsically solenoidal vector
$\B-\BRef$ that is \emph{tangent} to $\Surf$ and \emph{perpendicular} to
$\nhat$ on $\Surf$
\begin{equation}
\Helicity\rightarrow\Helicity+\int\limits_\Vol{d^3x}\,\grad\gauge\cdot\left(\B-\BRef\right)=\Helicity-\oint\limits_\Surf{dS}\,\nhat\cdot\left[\gauge\,\left(\B-\BRef\right)\right]=\Helicity.\label{eqn:Gauge:Invariant}  
\end{equation}    
All of the helicity terms developed in \S\ref{sec:Attribution} have the analogous
form and are similarly gauge invariant.\par
The potential magnetic field $\BPot$ is often used as a convenient
  reference field $\BRef$.  The potential field is harmonic
and thus admits a dual representation in terms of a vector potential or in
terms of the gradient of a scalar field
\begin{subequations}
\begin{equation}
\BRef\equiv\BPot=\grad\cross\APot=-\grad\MagPotential\qquad\x\in\Vol,\label{eqn:BPot}
\end{equation}
which satisfies  
\begin{align}
\grad\cross\BPot=\grad\cross\grad\cross\APot=-\grad\cross\grad\MagPotential=0&\qquad\x\in\Vol,\qquad(\mbox{No Currents})\label{eqn:NoCurrents}\\
\grad\cdot\BPot=\grad\cdot\grad\cross\APot=\nabla^2\MagPotential=0&\qquad\x\in\Vol.\qquad(\mbox{No Monopoles})\label{eqn:NoMonopoles}
\end{align}
A unique solution\footnote{This uniqueness in the Coulomb gauge is only
important for establishing a well-posed problem for determining $\APot$. Once
$\APot$ is determined, it may be gauge transformed without affecting
Equation~(\ref{eqn:HR}).} for the vector potential, $\APot$, requires an
arbitrary gauge condition for which the Coulomb gauge is a convenient choice
\cite[see Theorem 3.5 and Equations (3.23)-(3.25) in][]{Girault1986}
\begin{equation}
\grad\cdot\APot=0\quad\x\in\Vol\quad\mbox{and}\quad\nhat\cdot\APot=0\quad\x\in\Surf,\label{eqn:Neumann:Coulomb}
\end{equation}
and with this gauge choice, the vector potential satisfies the vector Poisson
equation
\begin{equation}
\nabla^2\APot=0\qquad\x\in\Vol,
\end{equation}
with boundary condition
\begin{equation}
\nhat\cdot\B=\nhat\cdot\grad\cross\APot=-\nhat\cdot\grad\MagPotential\qquad\x\in\Surf.\label{eqn:Neumann:f}
\end{equation}
\end{subequations}
Note that Equations~(\ref{eqn:NoMonopoles}) and~(\ref{eqn:Neumann:f}) define
the Neumann problem for the scalar potential $\MagPotential$ which is unique
to within an arbitrary scalar $\MagPotential_0$ and
Equations~(\ref{eqn:Neumann:Coulomb})\--(\ref{eqn:Neumann:f}) define a unique
vector potential for the same potential field $\BPot$.  The field $\BPot$ is
also the unique potential field that matches the normal component of $\B$ on
the boundary $\Surf$.  A reference field $\BRef$ that is potential is
`convenient' because no currents are supported by the potential field $\BPot$
in the volume of interest and thus the helicity of $\BRef=\BPot$ in a simply
connected domain may be intuitively defined as zero \cite[]{Berger1999}.
However, as noted in the Introduction (\S\ref{sec:Intro}), this convenience
comes at the price of possibly misrepresenting the origin of the fields.
Foreshadowing the development of \S\ref{sec:Attribution}, while $\BPot$
supports no internal currents in $\Vol$, Equation~(\ref{eqn:NoCurrents})
should not be interpreted to imply that $\BPot$ is \emph{produced} exclusively
by external currents!  For example see Figure~\ref{fig:Decomp}.  Furthermore,
while $\Bcl$ supports the internal currents $\JM$ in $\Vol$, it is generally
produced by both these internal currents and by external currents $\JME$ in
$\VolC$.  For example see Figure~\ref{fig:Closed}. This will be expanded on
further in \S\ref{sec:H:External}.  In particular, non-potential magnetic
fields in $\Vol$ can be produced by currents $\JME$ in the external universe
$\VolC$ when these currents thread the boundary $\Surf$ to enter and leave
$\Vol$.  We emphasize that $\B=\Bcl+\BPot$ is a \emph{mathematical}
decomposition determined by the geometry of the bounding surface $\Surf$ that
has no unique relationship with the origin of the magnetic field in
currents. \par
\cite{Berger1984a} showed that the evolution of the relative magnetic
helicity for an ideal plasma depends only on boundary terms that may
be computed from observables\footnote{Note that $\APot$ \emph{must} be
in the Coulomb gauge in Equation~(\ref{eqn:HF}) as the surface term is
not manifestly gauge invariant \cite[see][for an alternative formulation]{Schuck2019}.}
\begin{equation}
\frac{d \Helicity}{d t}=\frac{\partial}{\partial t}\int\limits_\Vol{d^3x}\left(\AV+\APot\right)\cdot\left(\B-\BPot\right)=-2\,c\,\oint\limits_\Surf{dS}\,\nhat\cdot\APotC\cross\E,  \label{eqn:HF}
\end{equation}
where $\grad\cdot\APotC=0$ in $\Vol$ is explicitly in the Coulomb gauge and
the electric field $\E=-\vel\cross\B/c$ is determined from the ideal Ohm's
law.  \cite{Berger1984b} further argued that this relative helicity
$\Helicity$ is a robust invariant for finite volumes such as those enclosing
flaring magnetic fields in the solar corona. A linear force-free field is the
absolute minimum energy state of a plasma in volume $\Vol$ with a prescribed
relative helicity $\Helicity$ and a specified or `line-tied' magnetic boundary
$\left.\nhat\cdot\B\right|_\Surf=g\left(\x\right)$ condition
\cite[]{Berger1984a,Berger1985,Jensen1984,Dixon1989,Laurence1991}.  \par
\cite{Berger1999,Berger2003} partitioned the relative helicity in
Equation~(\ref{eqn:HR}) into two further gauge invariant topological
quantities: the closed-closed helicity $\Hcl$
representing the linkages of magnetic field that closes in $\Vol$, and
the open-closed helicity $\HP$ representing the
linkages between the open magnetic field that threads the
boundary $\Surf$ and the magnetic field that closes inside $\Vol$:
\begin{equation}
\Helicity=\int\limits_\Vol{d^3x}\,\underbrace{\overbrace{\left(\AV-\APot\right)}^{\Acl}\cdot\overbrace{\left(\B-\BPot\right)}^{\Bcl}}_{\Hcl}+2\,\int\limits_\Vol{d^3x}\,\underbrace{\APot\cdot\overbrace{\left(\B-\BPot\right)}^{\Bcl}}_{\HP},\label{eqn:Berger2003}  
\end{equation}
where the Neumann potential magnetic field $\BPot$ has been
implemented as the reference field $\BRef$. Recently \cite{Pariat2017}
\cite[see also][]{Moraitis2014,Linan2018,Zuccarello2018} have
suggested that dynamic changes in the closed-closed 
  helicity $\Hcl$ may be a useful diagnostic of latent solar
  eruptivity leading to flares and coronal mass ejections.
  \cite{Pariat2017} denotes the first term in~(\ref{eqn:Berger2003})
  $\Helicity_J$ and designates it the ``current carrying helicity'' and
  denotes the second term $\Helicity_{\mathrm{P}J}$ and designates it the
  ``mutual helicity.'' \par
There is nothing physically special about the
linkages that close on themselves in $\Vol$ versus those that thread
the boundary $\Surf$: these fields are defined relative to a surface
$\Surf$ which often conveniently contains part of the photosphere in
solar observational investigations where magnetic field data is
regularly estimated from remote sensing observations, e.g.,
\textit{SDO}/HMI. In other words, $\BPot$ and $\Bcl$ are unique and
topologically distinct only in the context of the field $\B$
\emph{and} the volume $\Vol$ or equivalently $\Surf$. A different
volume $\Vol'$ bounded by a different surface $\Surf'$ will lead to
different potential fields $\BPot'\neq\BPot$ and magnetic fields
$\Bcl'\neq\Bcl$ that close in $\Vol'$ bounded by $\Surf'$, but
the same total $\B=\BPot+\Bcl=\BPot'+\Bcl'$, at the same location
$\x\in\Vol\cap\Vol'$. The local force $\J\cross\B$ on the plasma is
ultimately produced by the total magnetic field which contains no
information about the boundary $\Surf$, thus the magnetic field may be
decomposed in whatever way is convenient to identify the magnetic
topology and/or physical processes involved in the evolution of the
plasma. \par

\section{Relative  Helicity with Attribution for Magnetically `Open' Systems\label{sec:Attribution}}
\cite{Berger1984a} and \cite{Finn1985} established the relative helicity
$\Helicity$ as a gauge invariant measure of magnetic complexity in
magnetically open systems. The potential field $\BPot$ is a convenient
reference field as, \emph{mathematically}, the origin of $\BPot$ is in
currents supported by the external universe $\VolC$\----hence its helicity
$\int_\Vol{d^3x}\,\APot\cdot\BPot$ in $\Vol$ may be defined as zero in a
simply connected domain \cite[]{Berger1999}. However, as noted in
\cite{Schuck2022} and \S\ref{sec:Intro} and \S\ref{sec:Helmholtz} here, the
\emph{physical} origin of potential field may be in current supported in
$\Vol$. Thus the potential field $\BPot$ can misattribute the origin of flux
threading the bounding surface to external current sources. This insight
suggests that \citeauthor{Berger2003}'s decomposition of helicity in
Equation~(\ref{eqn:Berger2003}) may be further disentangled when the origin of
the magnetic field in currents is considered. The \cite{Berger1984a} and
\cite{Finn1985} formula~(\ref{eqn:HR}) is convenient for describing the
attribution of helicity because it is gauge agnostic\----we are free to write
$\AV$ and $\ARef$ in any gauge. Below in
\S\ref{sec:Internal}\--\ref{sec:Mutual} we extend \citeauthor{Berger2003}'s
decomposition of relative helicity in Equation~(\ref{eqn:Berger2003}) to
include attribution of the fields to their current sources in $\Vol$ and
$\VolC$. This motivates new definitions of internal and external and
  internal-external relative helicity distinguished by the domain of the
  current system that produces the magnetic linkages.  Our presentation is
general in that it is easily extended mutatis mutandis to a coronal volume
bounded by the photosphere and a boundary in the high corona or a box of
length $L$ on a side.
\subsection{Internal Relative Helicity in $\Vol$ Produced by Internal Sources: $\J$ in $\Vol$\label{sec:Internal}}
To compute the internal relative helicity produced by
internal sources we need to construct the field pairs
$\left(\AV,\B\right)$ and $\left(\APot,\BPot\right)$ produced by
\emph{internal} sources $\JM$. For the current system $\JM$ in the
domain of interest $\Vol$ the vector potential and magnetic field
follows directly from Equation~(\ref{eqn:BS:JM})
\begin{subequations}
\begin{align}
\AV\left(\J;t,\x\right)\equiv&\frac{1}{c}\,\int\limits_{\Vol}{d^3x'}\,\frac{\JM\left(t,\x'\right)}{\left|\x-\x'\right|}\qquad&\x\in\Rthree,\label{eqn:AV:I}\\
\bBS\left(\J;t,\x\right)\equiv&\grad\cross\AV\left(\J;t,\x\right)\qquad&\x\in\Rthree,\label{eqn:B:I}
\end{align}
\end{subequations}
which completely describes the attribution of the vector
potential and magnetic field produced by currents $\JM$ in
$\Vol$.\footnote{This is the most intuitive form for
  $\B\left(\JM;t,\x\right)$, but there may be more efficient
  techniques for computing it as described by
  Equations~(\ref{eqn:B:Surf:Externala})\--(\ref{eqn:B:Surf:External:Internal})
  in \S\ref{sec:Helmholtz}.} This field \emph{integrant} may be
decomposed in the usual fashion into magnetic fields that close in
$\Vol$ and magnetic fields that thread the boundary $\Surf$ using the
potential field methodology described above in
Equations~(\ref{eqn:BPot})\--(\ref{eqn:Neumann:f}). Explicitly this is
\begin{subequations}
\begin{align}
\BPot\left(\JM;t,\x\right)=&\grad\cross\APot\left(\JM;t,\x\right)\quad\mbox{and}\quad\BPot\left(\JM;t,\x\right)=-\grad\MagPotential\left(\JM;t,\x\right)\qquad&\x\in\Vol,\label{eqn:BPot:Internal}\\  
\grad\cdot\APot\left(\JM;t,\x\right)=&0\quad\x\in\Vol\quad\mbox{and}\quad\nhat\cdot\APot\left(\JM;t,\x\right)=0\quad&\x\in\Surf,\label{eqn:Neumann:Internal:Coulomb}\\
\nabla^2\APot\left(\JM;t,\x\right)=&0\quad\mbox{and}\quad\nabla^2\psi\left(\JM;t,\x\right)=0\qquad&\x\in\Vol,\\
\nhat\cdot\bBS\left(\JM;t,\x\right)=&\nhat\cdot\grad\cross\APot\left(\JM;t,\x\right)=-\nhat\cdot\grad\MagPotential\left(\JM;t,\x\right)\qquad&\x\in\Surf.\label{eqn:Neumann:Internal:f}
\end{align}
\end{subequations}
Note that
Equations~(\ref{eqn:BPot:Internal})\--(\ref{eqn:Neumann:Internal:f})
differ from~(\ref{eqn:BPot}), and
(\ref{eqn:Neumann:Coulomb})\--(\ref{eqn:Neumann:f}) in that the former
represents the potential field produced on the boundary by
\emph{physically} internal sources and
the latter represents the potential field produced on the boundary by
\emph{all physical} sources (\emph{internal and
external}).\footnote{Recall that the potential magnetic field
  $\BPot$ mathematically represents all current sources as external
  regardless of their physical origin. See for example, the discussion
  and Figure~\ref{fig:Decomp} in the Introduction
  (\S\ref{sec:Intro}).} This distinction is imposed by the boundary
conditions~(\ref{eqn:Neumann:Internal:f}) and~(\ref{eqn:Neumann:f}),
which in the former case is determined by the normal component of the
Biot-Savart law integrated over just the \emph{internal} sources $\J$
and in the latter case by the total field $\B$.\par
The magnetic fields that close in $\Vol$ and are produced by internal
current sources in $\Vol$ are then described as
\begin{subequations}
\begin{align}
  \Acl\left(\J;t,\x\right)\equiv&
  \AV\left(\J;t,\x\right)-\APot\left(\J;t,\x\right),\label{eqn:Acl:Internal}\\
  \Bcl\left(\J;t,\x\right)\equiv&\bBS\left(\J;t,\x\right)-\BPot\left(\J;t,x\right),\nonumber\\
  =&
  \grad\cross\Acl\left(\J;t,\x\right)=\grad\cross\left[\AV\left(\J;t,\x\right)-\APot\left(\J;t,\x\right)\right].\label{eqn:Bcl:Internal}
\end{align}  
\end{subequations}
The internal relative helicity which corresponds to the \emph{internal}
current sources is then
\begin{equation}
\Helicity\left(\JM,\JM\right)=\underbrace{\int_\Vol{d^3x}\,\overbrace{\left[\AV\left(\JM\right)-\APot\left(\JM\right)\right]}^{\Acl\left(\JM\right)}\cdot\overbrace{\left[\bBS\left(\JM\right)-\BPot\left(\JM\right)\right]}^{\Bcl\left(\JM\right)}}_{\Hcl\left(\JM,\JM\right)}+\underbrace{2\,\int_\Vol{d^3x}\,\APot\left(\JM\right)\cdot\overbrace{\left[\bBS\left(\JM\right)-\BPot\left(\JM\right)\right]}^{\Bcl\left(\JM\right)}}_{\HP\left(\JM,\JM\right)},\label{eqn:H:I}  
\end{equation}
where the independent variables $t$ and $\x$ have been suppressed for
brevity. Both integrals are gauge invariant because
$\grad\cdot\Bcl\left(\JM\right)=0$ and
$\nhat\cdot\left.\Bcl\left(\JM\right)\right|_\Surf=0$ by construction (see
Equation~(\ref{eqn:Gauge:Invariant})). In the \cite{Berger1999,Berger2003}
paradigm, this expression describes the closed-closed helicity
$\Hcl\left(\JM,\JM\right)$ of the magnetic field that is produced by internal
current sources and closes in $\Vol$ and the open-closed helicity
$\HP\left(\JM,\JM\right)$ between the magnetic field that is produced by
internal current sources and closes in $\Vol$ and the magnetic field that is
produced by internal current sources and threads the boundary
$\Surf$. However, in our new paradigm $\Helicity\left(\JM,\JM\right)$
represents the total internal relative helicity in $\Vol$ of magnetic field
produced by currents $\JM$ in $\Vol$. This is arguably the true self-helicity
of the current system $\JM$ in $\Vol$.  If there were no external currents
$\JME$, then Equations~(\ref{eqn:Berger2003}) and~(\ref{eqn:H:I}) would
produce identical values.
\subsection{External Relative Helicity in $\Vol$ Produced by External Sources: $\JME$ in $\VolC$\label{sec:H:External}}
To compute the external relative helicity produced by external
sources we need to construct the field pairs $\left(\AV,\B\right)$ and
$\left(\APot,\BPot\right)$ produced by \emph{external} sources $\JME$.  The
magnetic vector potential $\AV$ and corresponding magnetic field $\B$ produced
by external current sources follows directly from (\ref{eqn:BS:JME})
\begin{subequations}
\begin{align}
\AV\left(\JME;t,\x\right)\equiv&\frac{1}{c}\,\int\limits_{\VolC}{d^3x'}\,\frac{\JME\left(t,\x'\right)}{\left|\x-\x'\right|}\qquad&\x\in\Rthree,\label{eqn:AV:E}\\
\bBS\left(\JME;t,\x\right)\equiv&\grad\cross\AV\left(\JME;t,\x\right)\qquad&\x\in\Rthree,\label{eqn:B:E}
\end{align}
\end{subequations}
where the domain of integration is over the entire external volume
$\VolC$ that contains current sources $\JME$. However, in practice, we
do not have access to this information. Usually, at \emph{best}, we
have information about the currents $\JM$ in our domain of interest
$\Vol$ and information on the boundary $\Surf$ and so while
Equations~(\ref{eqn:AV:E})\--(\ref{eqn:B:E}) are formally correct and
useful for developing insight, they are not practical for
computation. However, the magnetic field due to external sources can
be computed from
(\ref{eqn:B:Surf:Externala})\--(\ref{eqn:B:Surf:External}). We
emphasize again here that determining $\bBS\left(\JME;t,\x\right)$ in
$\Vol$ does not require performing the Biot-Savart integral over
$\JME$ in $\VolC$. \par
Again, this field may be decomposed in the usual fashion into magnetic
fields that close in $\Vol$ and magnetic fields that thread the
boundary $\Surf$ using the potential field methodology described
above in
equations~(\ref{eqn:BPot})\--(\ref{eqn:Neumann:f}). Explicitly this is
\begin{subequations}
\begin{align}
\BPot\left(\JME;t,\x\right)=&\grad\cross\APot\left(\JME;t,\x\right)\quad\mbox{and}\quad\BPot\left(\JME;t,\x\right)=-\grad\MagPotential\left(\JME;t,\x\right)\qquad&\x\in\Vol,\label{eqn:BPot:External}\\  
\grad\cdot\APot\left(\JME;t,\x\right)=&0\quad\x\in\Vol\quad\mbox{and}\quad\nhat\cdot\APot\left(\JME;t,\x\right)=0\quad&\x\in\Surf,\label{eqn:Neumann:External:Coulomb}\\
\nabla^2\APot\left(\JME;t,\x\right)=&0\quad\mbox{and}\quad\nabla^2\psi\left(\JME;t,\x\right)=0\qquad&\x\in\Vol,\\
\nhat\cdot\bBS\left(\JME;t,\x\right)=&\nhat\cdot\grad\cross\APot\left(\JME;t,\x\right)=-\nhat\cdot\grad\MagPotential\left(\JME;t,\x\right)\qquad&\x\in\Surf.\label{eqn:Neumann:External:f}
\end{align}
\end{subequations}
Combining the results in Equations~(\ref{eqn:BPot}),~(\ref{eqn:BPot:Internal})
and~(\ref{eqn:BPot:External}), and making use of
Equations~(\ref{eqn:Neumann:f}),~(\ref{eqn:Neumann:Internal:f}),~(\ref{eqn:Neumann:External:f}),
and~(\ref{eqn:B:All}),
\begin{equation}
\MagPotential\left(t,\x\right)=\MagPotential\left(\JM;t,\x\right)+\MagPotential\left(\JME;t,\x\right)\qquad\x\in\Vol, 
\end{equation}
or
\begin{equation}
\BPot\left(t,\x\right)=\BPot\left(\JM;t,\x\right)+\BPot\left(\JME;t,\x\right)\qquad\x\in\Vol,\label{eqn:P:All}
\end{equation}
and we see that the traditional Neumann potential field described by
$\BPot=-\grad\psi$ conflates the magnetic field produced by internal current
sources $\JM$ and external current sources $\JME$ as discussed in the
introduction. A similar conflation occurs for the closed field $\Bcl$. The
closed field produced by external currents is
\begin{equation}
\Bcl\left(\JME;t,x\right)=\bBS\left(\JME;t,\x\right)-\BPot\left(\JME;t,\x\right)\qquad\x\in\Vol,\label{eqn:Bcl:External}
\end{equation}
and then combining the results in Equations~(\ref{eqn:Bcl}),~
(\ref{eqn:Bcl:Internal}) and~(\ref{eqn:Bcl:External}), and making use of 
Equations~(\ref{eqn:BPot}),~(\ref{eqn:B:All}), and~(\ref{eqn:P:All}),
\begin{equation}
\Bcl\left(t,x\right)=\Bcl\left(\JM;t,x\right)+\Bcl\left(\JME;t,x\right)=\B\left(t,x\right)-\BPot\left(t,x\right)\qquad\x\in\Vol.\label{eqn:Bcl:Both}
\end{equation}\par
For the external current system $\JM$ to \emph{contribute} to the closed field
$\Bcl$ in $\Vol$ is perhaps not surprising given
Figure~\ref{fig:Closed}. However, the external current system $\JME$ in
$\VolC$ can produce closed-closed helicity $\Hcl$ in $\Vol$ by generating
closed field in $\Vol$ on its own!  The presence of current systems that pass
from $\Vol$ to $\VolC$ or vice versa also implies that
\begin{equation}
\grad\cross\Bcl\left(\JME;t,\x\right)\neq0\qquad\x\in\Vol.
\end{equation}  
\emph{The external current, $\JME$, in volume, $\VolC$, injects magnetic
vorticity into $\Vol$.} This is apparent if we consider the curl
of~(\ref{eqn:B:E})
\begin{equation}
\grad\cross\bBS\left(\JME;t,\x\right)=\frac{1}{c}\,\grad\cross\grad\cross\int\limits_{\VolC}{d^3x'}\,\frac{\JME\left(t,\x'\right)}{\left|\x-\x'\right|}\qquad\x\in\Rthree,  
\end{equation}  
where the observation point $\x$ is in $\Vol$ not $\VolC$. Using the vector
relationship
\begin{equation}  
\grad\cross\grad\cross\boldsymbol{a}=\grad\left(\grad\cdot\boldsymbol{a}\right)-\nabla^2\boldsymbol{a},
\end{equation}  
this becomes 
\begin{equation}
\grad\cross\bBS\left(\JME;t,\x\right)=\frac{1}{c}\,\grad\left[\grad\cdot\int\limits_{\VolC}{d^3x'}\,\frac{\JME\left(t,\x'\right)}{\left|\x-\x'\right|}\right]-\frac{1}{c}\,\nabla^2\int\limits_{\VolC}{d^3x'}\,\frac{\JME\left(t,\x'\right)}{\left|\x-\x'\right|}\qquad\x\in\Rthree.  
\end{equation}  
The kernel in the second term has the form of a delta distribution
  because
\begin{equation}
  \nabla^2\left|\x-\x'\right|^{-1}=-4\,\pi\,\left\lbrace\begin{array}{lr}
  \delta\left(\x-\x'\right)&\x\in\VolC,\\
  \FactorC\left(x\right)\,\delta\left(\x-\x'\right)&\x\in\SurfC,\\
  0&\x\in\Vol.
  \end{array}\right.
\end{equation}
leading to
\begin{equation}
  \grad\cross\bBS\left(\JME;t,\x\right)-\frac{1}{c}\,\grad\left[\grad\cdot\int\limits_{\VolC}{d^3x'}\,\frac{\JME\left(t,\x'\right)}{\left|\x-\x'\right|}\right]=\frac{4\,\pi}{c}\,\FactorC\left(\x\right)\,\JME\left(t,\x\right)\qquad\x\in\Rthree,
\end{equation}
where
\begin{equation}
  \FactorC\left(\x\right)=1-\alpha\left(\x\right)=\left\lbrace\begin{array}{ll}
  1\qquad&\x\in\VolC\\
\left. \begin{array}{ll}1/2&\mbox{smooth surfaces}\\
    3/4&\mbox{edges of $\Vol$}\\
  7/8&\mbox{vertices of $\Vol$}\end{array}\right\rbrace\qquad&\x\in\SurfC\\
  0\qquad&\x\in\Vol
\end{array}\right.,\label{eqn:AlphaS}  
\end{equation}
follows from Equation~(\ref{eqn:Alpha}) for $\Factor\left(\x\right)$ and for
the values in braces we have assumed that $\VolC$ encloses $\Vol$.
Passing the divergence under the integral operator, using
\begin{equation}
  \grad\frac{1}{\left|\x-\x'\right|}=-\grad'\frac{1}{\left|\x-\x'\right|},
\end{equation}
 and
\begin{equation}
\grad\cdot\left(\phi\,\boldsymbol{a}\right)=\boldsymbol{a}\cdot\grad\phi+\phi\,\grad\cdot\boldsymbol{a},\label{eqn:Div_psi_f}  
\end{equation}
this simplifies to 
\begin{equation}
  \grad\cross\bBS\left(\JME;t,\x\right)+\frac{1}{c}\,\grad\left[\int\limits_{\VolC}{d^3x'}\,\grad'\cdot\,\frac{\JME\left(t,\x'\right)}{\left|\x-\x'\right|}-\int\limits_{\VolC}{d^3x'}\,\frac{\grad'\cdot\JME\left(t,\x'\right)}{\left|\x-\x'\right|}\right]=\frac{4\,\pi}{c}\,\FactorC\left(\x\right)\,\JME\left(t,\x\right)\qquad\x\in\Rthree.\label{eqn:Simplifies}  
\end{equation}  
The Gauss-Ostrogradsky theorem where $\nhat$ points into $\Vol$
\begin{equation}
\int\limits_{\Vol}{d^3x}\,\grad\cdot\boldsymbol{a}=-\oint\limits_{\Surf}{dS}\,\nhat\cdot\boldsymbol{a},\label{eqn:Gauss}
\end{equation}  
relates the volume integral of the divergence to a surface integral over the
normal component at the boundary of the volume.  Then with the
  Gauss-Ostrogradsky theorem, Equation~(\ref{eqn:Simplifies}) becomes
\begin{subequations}  
\begin{equation}
  \grad\cross\bBS\left(\JME;t,\x\right)-\frac{1}{c}\,\overbrace{\grad\left[\oint\limits_{\SurfC}{dS'}\,\frac{\nhat'\cdot\JME\left(t,\x'\right)}{\left|\x-\x'\right|}+\int\limits_{\VolC}{d^3x'}\,\frac{\grad'\cdot\JME\left(t,\x'\right)}{\left|\x-\x'\right|}\right]}^{{\partial\E\left(\JME;t,\x\right)}/{\partial
  t}}=\frac{4\,\pi}{c}\,\FactorC\left(\x\right)\,\JME\left(t,\x\right)\qquad\x\in\Rthree,\label{eqn:Curl:bBS:JME:a}  
\end{equation}
or with $\grad\cdot\JME=0$ for $\x\in\Vol$
\begin{equation}
  \grad\cross\bBS\left(\JME;t,\x\right)+\frac{1}{c}\,\overbrace{\grad\oint\limits_{\Surf}{dS'}\,\frac{\nhat'\cdot\JM\left(t,\x'\right)}{\left|\x-\x'\right|}}^{-{\partial\E\left(\JME;t,\x\right)}/{\partial
  t}}=0\qquad\x\in\Vol.\label{eqn:Curl:bBS:JME}  
\end{equation}
where in the last expression we have taken the surface integral with
  respect to $\Surf$ instead of $\SurfC$, used $\grad\cdot\JME=0$ as implied
  by Ampere's law~(\ref{eqn:Ampere}), and assumed that the normal component
of the current is continuous across the boundary
$\nhat\cdot\left(\JM-\JME\right)=0$ for
$\x\in\Surf$. Equation~(\ref{eqn:Curl:bBS:JME}) has the form of the
Amp{\`e}re-Maxwell equation
\begin{equation}
\grad\cross\bBS\left(\JME;t,\x\right)-\frac{1}{c}\,\frac{\partial\E\left(\JME;t,\x\right)}{\partial
  t}=0\qquad\x\in\Vol,
\end{equation}
\end{subequations}  
where there is no external material current $\JME$ in $\Vol$. Thus, the
magnetic vorticity produced in $\Vol$ is \emph{balanced, but not generated},
by a time-dependent electric field in $\Vol$, the so-called `displacement
current,' and both are produced by $\JME$ in $\VolC$ or on $\Surf$. We
emphasize that the displacement current $\partial \E/\partial t$ \emph{is not
a source of magnetic field.}  Closed magnetic field in the sense of
$\oint\,\B\left(\JME;t,\x\right)\cdot{d\ell}\neq0$ for $\x\in\Vol$, indicates
the presence of magnetic vorticity\----and not the exclusive presence of a
local material current $\JM$. The source of this magnetic vorticity in $\Vol$
may be a non-local current source, e.g., $\JME$ in $\VolC$.\par
The presence of displacement currents must be reconciled with
  Amp{\'e}re's law~(\ref{eqn:Ampere}).  Consider
$\grad\cross\bBS\left(\JM;t,\x\right)$ by following the derivation
of~(\ref{eqn:Curl:bBS:JME:a}) mutatis mutandis
\begin{subequations}  
\begin{equation}
  \grad\cross\bBS\left(\JM;t,\x\right)-\frac{1}{c}\,\overbrace{\grad\left[\oint\limits_{\Surf}{dS'}\,\nhat'\cdot\,\frac{\JM\left(t,\x'\right)}{\left|\x-\x'\right|}+\int\limits_{\Vol}{d^3x'}\,\frac{\grad'\cdot\JM\left(t,\x'\right)}{\left|\x-\x'\right|}\right]}^{{\partial\E\left(\JM;t,\x\right)}/{\partial
  t}}=\frac{4\,\pi}{c}\,\Factor\left(\x\right)\,\JM\left(t,\x\right)\qquad\x\in\Rthree,\label{eqn:Curl:bBS:JM:a}
  \end{equation}
or with $\grad\cdot\J=0$ for $\x\in\Vol$
\begin{equation}
\grad\cross\bBS\left(\JM;t,\x\right)-\frac{1}{c}\,\overbrace{\grad\oint\limits_{\Surf}{dS'}\,\frac{\nhat'\cdot\JM\left(t,\x'\right)}{\left|\x-\x'\right|}}^{{\partial\E\left(\JM;t,\x\right)}/{\partial
  t}}=\frac{4\,\pi}{c}\,\J\left(t,\x\right)\qquad\x\in\Vol.\label{eqn:Curl:bBS:JM}  
\end{equation}
which has a material current because
$\nabla^2\left|\x-\x'\right|^{-1}=-4\,\pi\,\delta\left(\x-\x'\right)$ is a
delta distribution for
$\x\in\Vol$ and $\x'\in\Vol$. 
This also has the form of the Amp{\`e}re-Maxwell equation
\begin{equation}  
\grad\cross\bBS\left(\JM;t,\x\right)-\frac{1}{c}\,\frac{\partial\E\left(\JM;t,\x\right)}{\partial
  t}=\frac{4\,\pi}{c}\,\JM\left(t,\x\right)\qquad\x\in\Vol.
\end{equation}  
\end{subequations}  
\begin{subequations} 
Combining Equations~(\ref{eqn:Curl:bBS:JME:a}) and~(\ref{eqn:Curl:bBS:JM:a}) 
\begin{align}
\grad\cross\bBS\left(\JM;t,\x\right)+\grad\cross\bBS\left(\JME;t,\x\right)-&\frac{1}{c}\,\overbrace{\left[\grad\oint\limits_{\Surf}{dS'}\,\frac{\nhat'\cdot\JM\left(t,\x'\right)}{\left|\x-\x'\right|}-\grad\oint\limits_{\Surf}{dS'}\,\frac{\nhat'\cdot\JM\left(t,\x'\right)}{\left|\x-\x'\right|}\right]}^{\mbox{Net
    Displacement Current}}=
\nonumber\\ &\qquad\qquad\qquad\qquad\frac{4\,\pi}{c}\,\left[\Factor\left(\x\right)\,\JM\left(t,\x\right)+\FactorC\left(\x\right)\,\JME\left(t,\x\right)\right]\qquad\x\in\Rthree,
\end{align}
the displacement currents cancel and 
\begin{align}
\grad\cross\overbrace{\left[\bBS\left(\JM;t,\x\right)+\bBS\left(\JME;t,\x\right)\right]}^{\B}=\frac{4\,\pi}{c}\,\jm\qquad\x\in\Rthree,\\
\jm\left(t;\x\right)=\alpha\left(\x\right)\,\JM\left(t,\x\right)+\left[1-\alpha\left(\x\right)\right]\,\JME\left(t,\x\right)\qquad\x\in\Rthree
\end{align}
\end{subequations}
recovers Amp{\'e}re's law~(\ref{eqn:Ampere}) with $\grad\cdot\jm=0$ for
$\x\in\Rthree$. Thus, even when the \emph{net displacement current density is
zero in $\Vol$,} as implied by Amp{\`e}re's law~(\ref{eqn:Ampere}), there may
be external contributions from $\JME$ in $\VolC$ to $\Bcl$ and displacement
currents in $\Vol$. \par
To summarize the results to this point, determining the helicities due to
internal sources requires computation of the vector potential
$\AV\left(\JM;t,\x\right)$ and magnetic field
$\bBS\left(\JM;t,\x\right)$ via the Biot-Savart
law~(\ref{eqn:AV:I})\--(\ref{eqn:B:I}). The magnetic field produced
by external sources $\bBS\left(\JME;t,\x\right)$ may be computed by
subtracting $\bBS\left(\JM;t,\x\right)$ from the total field $\B$ as
in Equation~(\ref{eqn:B:Surf:Externala}).  The decomposition of these
attributed fields into components that close in $\Vol$ and that thread
the boundary $\Surf$ requires constructing field pairs
$\APot\left(\J;t,\x\right)$ and $\BPot\left(\J;t,\x\right)$ and
$\APot\left(\JME;t,\x\right)$ and $\BPot\left(\JME;t,\x\right)$ in
Equations~(\ref{eqn:BPot:Internal})\--(\ref{eqn:Neumann:Internal:f}) and Equations~(\ref{eqn:BPot:External})\--(\ref{eqn:Neumann:External:f})
which in turn may be used to construct
$\Acl\left(\JM;t,\x\right)$, $\Bcl\left(\JM;t,\x\right)$ and
$\Bcl\left(\JME;t,\x\right)$ in Equations~(\ref{eqn:Acl:Internal}), (\ref{eqn:Bcl:Internal}), and
  (\ref{eqn:Bcl:External}).\par
The last missing piece is to compute $\Acl\left(\JME;t,\x\right)$ from
what we already know. First recall that
\begin{equation}
\grad\cdot\Bcl\left(\JME;t,\x\right)=0\quad\x\in\Vol\quad\mbox{and}\quad\nhat\cdot\Bcl\left(\JME;t,\x\right)=0\quad\x\in\Surf,
\end{equation}
is an \emph{intrinsically solenoidal} vector field. Thus, $\Acl\left(\JME;t,\x\right)$ may be
  reconstructed in the Coulomb gauge with the Biot-Savart operator
\begin{equation}
\Acl\left(\JME;t,\x\right)=\frac{1}{4\,\pi}\,\grad\cross\int\limits_\Vol{d^3x'}\,\frac{\Bcl\left(\JME;t,\x\right)}{\left|\x-\x'\right|}\qquad\x\in\Vol\cup\Surf.\label{eqn:Acl:Simple}
\end{equation}
This is perhaps the conceptually simplest expression for
$\Acl\left(\JME;t,\x\right)$, but alternatives are presented in
Appendix~\ref{sec:AV:JME}.  The external relative helicity which
  corresponds to the \emph{external} current sources is then
\begin{equation}
\Helicity\left(\JME,\JME\right)=\underbrace{\int_\Vol{d^3x}\,\overbrace{\left[\AV\left(\JME\right)-\APot\left(\JME\right)\right]}^{\Acl\left(\JME\right)}\cdot\overbrace{\left[\bBS\left(\JME\right)-\BPot\left(\JME\right)\right]}^{\Bcl\left(\JME\right)}}_{\Hcl\left(\JME,\JME\right)}+\underbrace{2\,\int_\Vol{d^3x}\,\APot\left(\JME\right)\cdot\overbrace{\left[\bBS\left(\JME\right)-\BPot\left(\JME\right)\right]}^{\Bcl\left(\JME\right)}}_{\HP\left(\JME,\JME\right)},\label{eqn:H:E}  
\end{equation}
where we have again dropped the temporal and spatial variables for
convenience. In the \cite{Berger1999,Berger2003} paradigm, this expression
describes the closed-closed helicity $\Hcl\left(\JME,\JME\right)$ of the
magnetic field that is produced by external current sources and closes in
$\Vol$ and the open-closed helicity $\HP\left(\JME,\JME\right)$ between
magnetic field that is produced by external current sources and closes in
$\Vol$ and the magnetic field that is produced by external current sources and
threads the boundary $\Surf$. However, in our new paradigm
$\Helicity\left(\JME,\JME\right)$ represents the total external relative
helicity in $\Vol$ of magnetic field produced by currents $\JME$ in
$\VolC$. This is arguably the true self-helicity of the current system $\JME$
in $\Vol$. If there were no internal currents $\JM$, then
Equations~(\ref{eqn:Berger2003}) and~(\ref{eqn:H:E}) would produce identical
values.
\subsection{The Relative Helicity of the Mutual Linkages Between the Internal $\JM$  and External $\JME$ Sources \label{sec:Mutual}}
Above we have established four gauge invariant quantities that describe the
relative helicity of the linkages produced by currents $\JM$ and
$\JME$ in $\Vol$ and $\VolC$, respectively: $\Hcl\left(\JM,\JM\right)$,
$\HP\left(\JM,\JM\right)$, $\Hcl\left(\JME,\JME\right)$, and
$\HP\left(\JME,\JME\right)$. Four other gauge invariant quantities may be
constructed that describe the relative helicity of the mutual linkages between
fields that have their origin in currents $\JM$ and $\JME$ in the internal
$\Vol$ and external $\VolC$ volumes, respectively:
\begin{subequations}
\begin{align}
  \Helicity\left(\JM,\JME\right)=&\underbrace{\int_\Vol{d^3x}\,\overbrace{\left[\AV\left(\JM\right)-\APot\left(\JM\right)\right]}^{\Acl\left(\JM\right)}\cdot\overbrace{\left[\bBS\left(\JME\right)-\BPot\left(\JME\right)\right]}^{\Bcl\left(\JME\right)}}_{\Hcl\left(\JM,\JME\right)}+
  \underbrace{2\,\int_\Vol{d^3x}\,\APot\left(\JM\right)\cdot\overbrace{\left[\bBS\left(\JME\right)-\BPot\left(\JME\right)\right]}^{\Bcl\left(\JME\right)}}_{\HP\left(\JM,\JME\right)},\label{eqn:H:IE}\\  
\Helicity\left(\JME,\JM\right)=&\underbrace{\int_\Vol{d^3x}\,\overbrace{\left[\AV\left(\JME\right)-\APot\left(\JME\right)\right]}^{\Acl\left(\JME\right)}\cdot\overbrace{\left[\bBS\left(\JM\right)-\BPot\left(\JM\right)\right]}^{\Bcl\left(\JM\right)}}_{\Hcl\left(\JME,\JM\right)}+\underbrace{2\,\int_\Vol{d^3x}\,\APot\left(\JME\right)\cdot\overbrace{\left[\bBS\left(\JM\right)-\BPot\left(\JM\right)\right]}^{\Bcl\left(\JM\right)}}_{\HP\left(\JME,\JM\right)}.\label{eqn:H:EI} 
\end{align}
\end{subequations}
\Referee{Note that $\Hcl\left(\JM,\JME\right)=\Hcl\left(\JME,\JM\right)$ by
  reciprocity, but
  $\Helicity\left(\JM,\JME\right)\neq\Helicity\left(\JME,\JM\right)$ because
  $\HP\left(\JM,\JME\right)\neq\HP\left(\JME,\JM\right)$. To prove
  reciprocity, the difference between the integrands of the first term in
  Equations~(\ref{eqn:H:IE}) and~(\ref{eqn:H:EI}) may be expressed
\begin{subequations}  
\begin{align}  
\Acl\left(\JM\right)\cdot\Bcl\left(\JME\right)-\Acl\left(\JME\right)\cdot\Bcl\left(\JM\right)
=&\Acl\left(\JM\right)\cdot\grad\cross\Acl\left(\JME\right)-\grad\cross\Acl\left(\JM\right)\cdot\Acl\left(\JME\right),\\
=&\grad\cdot\left[\Acl\left(\JME\right)\cross\Acl\left(\JM\right)\right].\label{eqn:this}
\end{align}  
\end{subequations}
Recalling that $\Hcl$ is gauge invariant and a vector potential that produces
closed magnetic field on $\Surf$ may be expressed $\Acl=A\,\nhat+\grad\Lambda$
for $\x\in\Surf$, the integral of Equation~(\ref{eqn:this}) with
Equation~(\ref{eqn:Gauss}) becomes
\begin{equation}
\Hcl\left(\JM,\JME\right)-\Hcl\left(\JME,\JM\right)=-\oint\limits_\Surf{dS}\,\nhat\cdot\left[\Acl\left(\JME\right)\cross\Acl\left(\JM\right)\right]=0.
\end{equation}}
\subsection{Discussion}
The superposition of
Equations~(\ref{eqn:H:I}),~(\ref{eqn:H:E}),~(\ref{eqn:H:IE}),
and~(\ref{eqn:H:EI}) reconstructs the relative helicity in
Equation~(\ref{eqn:HR}), for $\BRef=\BPot$. Thus the traditional relative
helicity may be decomposed into eight gauge invariant quantities that describe
both the self-linking of magnetic field that closes in $\Vol$ and the mutual
linking between magnetic field that closes in $\Vol$ and magnetic field that
threads the boundary, while simultaneously distinguishing the physical origin
of the magnetic field with currents $\JM$ in $\Vol$ and $\JME$ in $\VolC$. In
the \cite{Berger1999,Berger2003} paradigm these eight terms are arranged into
`self' and `mutual' helicity \emph{based on the magnetic field properties
open, $\BPot$, or closed, $\Bcl$, on the boundary $\Surf$,} regardless of
their origin in currents $\JM$ and $\JME$ in $\Vol$ and $\VolC$:
\begin{equation}
  \Helicity=\underbrace{\Hcl\left(\JM,\JM\right)+\Hcl\left(\JME,\JME\right)+\Hcl\left(\JM,\JME\right)+\Hcl\left(\JME,\JM\right)}_{\mbox{closed-closed}~\Hcl~\mbox{(`self')}}+
  \underbrace{\HP\left(\JM,\JM\right)+\HP\left(\JME,\JME\right)+\HP\left(\JM,\JME\right)+
  \HP\left(\JME,\JM\right)}_{\mbox{open-closed}~\HP~\mbox{(`mutual')}}.
\end{equation}\par
In our new paradigm, these terms are arranged into internal or external (self)
and internal-external (mutual) helicity \emph{based on their origin in
currents $\JM$ and $\JME$ in $\Vol$ and $\VolC$, respectively,} regardless of
the magnetic field properties on the boundary $\Surf$:
\begin{subequations}
\begin{align}
  \Helicity=&\overbrace{\Helicity\left(\JM,\JM\right)}^{{\mbox{internal}}\atop{\mbox{(self)}}}+\overbrace{\Helicity\left(\JME,\JME\right)}^{{\mbox{external}}\atop{\mbox{(self)}}}+\overbrace{\Helicity\left(\JM,\JME\right)+\Helicity\left(\JME,\JM\right)}^{\mbox{internal-external (mutual)}},\\
  =&\underbrace{\Hcl\left(\JM,\JM\right)+\HP\left(\JM,\JM\right)}_{\mbox{internal (self)}}+
  \underbrace{\Hcl\left(\JME,\JME\right)+\HP\left(\JME,\JME\right)}_{\mbox{external (self)}}+
  \underbrace{\Hcl\left(\JM,\JME\right)+\HP\left(\JM,\JME\right)+
  \Hcl\left(\JME,\JM\right)+\HP\left(\JME,\JM\right)}_{\mbox{internal-external (mutual)}}.
\end{align}
\end{subequations}
 \Referee{Seven of these components are independent as
   $\Hcl\left(\JME,\JM\right)=\Hcl\left(\JM,\JME\right)$.}  We emphasize that
 each of the \Referee{seven independent} components of relative helicity in
 this decomposition is gauge invariant, in isolation, a quality of a valid
 observable also emphasized recently by \cite{Schuck2019}.  This more
 comprehensive set of \Referee{seven} helicity components provides a basis for
 a more detailed examination of the interplay between internally and
 externally sourced magnetic fields involved in reconnection during solar
 eruptions and potentially reconnection in the tail and magnetopause during
 terrestrial geomagnetic storms. \par

\section{The Magnetic Energy\label{sec:Energy}}
As described above, the magnetic field in $\Vol$ may be
decomposed with the magnetic field components that simultaneously distinguish
their physical origin as
\begin{equation}
\B=\BPot\left(\JM\right)+\Bcl\left(\JM\right)+\BPot\left(\JME\right)+\Bcl\left(\JME\right).\label{eqn:B:Energy:All}  
\end{equation}
The \emph{local} magnetic energy density is then comprised of 10 distinct
terms proportional to:
\begin{align}
  B^2=&\BPot^2\left(\JM\right)+\BPot^2\left(\JME\right)+2\,\BPot\left(\JM\right)\cdot\BPot\left(\JME\right)\nonumber\\
  &+\Bcl^2\left(\JM\right)+\Bcl^2\left(\JME\right)+2\,\Bcl\left(\JM\right)\cdot\Bcl\left(\JME\right)\nonumber\\
  &+2\,\Bcl\left(\JM\right)\cdot\left[\BPot\left(\JM\right)+\BPot\left(\JME\right)\right]+2\,\Bcl\left(\JME\right)\cdot\left[\BPot\left(\JM\right)+\BPot\left(\JME\right)\right].\label{eqn:E:Density}
\end{align}
The first row involves exclusively the energy density of magnetic field
that threads the boundary $\Surf$. The second row of terms involves
exclusively the energy density of magnetic field that closes in $\Vol$.
The bottom row describes the mutual energy density between magnetic field
that threads the boundary $\Surf$ and the magnetic field that closes in
$\Vol$. The magnetic energy is
\begin{equation}
\Energy\equiv\frac{1}{8\,\pi}\int\limits_\Vol{d^3x}\,{B^2}.
\end{equation}
Note that mathematically both
$\BPot\left(\JM\right)=-\grad\psi\left(\JM\right)$ and
$\BPot\left(\JME\right)=-\grad\psi\left(\JME\right)$ may be described as the
gradient of a scalar in the volume of interest $\Vol$. Thus, the bottom row of
terms in~(\ref{eqn:E:Density}) does not contribute to the net magnetic energy
in $\Vol$ as with identities~(\ref{eqn:Div_psi_f}) and~(\ref{eqn:Gauss})
\begin{equation}
  \int\limits_\Vol{d^3x}\,\Bcl\cdot\grad\psi=\int\limits_\Vol{d^3x}\,\grad\cdot\left(\psi\,\Bcl\right)=-\oint_\Surf{dS}\,\psi\,\nhat\cdot\Bcl=0,
\end{equation}
resulting in
\begin{align}
  \Energy=&\overbrace{-\frac{1}{8\,\pi}\int\limits_\Vol{d^3x}\,\left[\BPot\left(\JM\right)\cdot\grad\MagPotential\left(\JM\right)+\BPot\left(\JME\right)\cdot\grad\MagPotential\left(\JME\right)
 +2\,\BPot\left(\JM\right)\cdot\grad\MagPotential\left(\JME\right)\right]}^{\Energy_\mathrm{Potential}}\nonumber\\
&+\underbrace{\frac{1}{8\,\pi}\,\int\limits_\Vol{d^3x}\,\left[\Bcl^2\left(\JM\right)+\Bcl^2\left(\JME\right)+2\,\Bcl\left(\JM\right)\cdot\Bcl\left(\JME\right)\right]}_{\Energy_\mathrm{Free}}.\label{eqn:EMF}
\end{align}
\subsection{The Pre and Post-Eruptive State of the Corona: Is the `Free Energy' Relevant?}
The potential field is a useful reference field for solar eruptions
because only small changes in the normal component of the magnetic field are
observed when comparing pre and post solar eruptions \cite[]{Wang1992a,Wang1994,Sudol2005,Wang2006,Sun2017}. Thus,
the potential field $\BPot$ is believed to remain constant during the
eruption.  The potential state $\Energy_{\mathrm{Potential}}$ with $\BPot$
matching $\nhat\cdot\B$ on $\Surf$ is often proven to be the `minimum energy
state' for volume $\Vol$ \cite[see for example][]{Priest2014}.
Consequently, the maximum `free energy' $\Energy_\mathrm{Free}$ of the
corona that is available to drive solar eruptions while holding that normal
component fixed in the photosphere has been computed as the difference
between the energy of the magnetic field, $\Energy$, in the coronal volume $\Vol$
and the energy of this potential magnetic field, $\Energy_\mathrm{Potential}$,
where
\cite[]{Tanaka1973,Yang1983a,Gary1987,Sakurai1987,Low1990,Klimchuk1992,Tarr2013,Zhang2016,Schuck2019,Liu2023}
\begin{subequations}
\begin{equation}
\Energy_{\mathrm{Potential}}=-\frac{1}{8\,\pi}\int\limits_\Vol{d^3x}\,\left[\BPot\left(\JM\right)\cdot\grad\MagPotential\left(\JM\right)+\BPot\left(\JME\right)\cdot\grad\MagPotential\left(\JME\right)
 +2\,\BPot\left(\JM\right)\cdot\grad\MagPotential\left(\JME\right)\right],\label{eqn:EP}
\end{equation}
and
\begin{equation}
\Energy_{\mathrm{Free}}=\Energy-\Energy_{\mathrm{Potential}}=\frac{1}{8\,\pi}\,\int\limits_\Vol{d^3x}\,\left[\Bcl^2\left(\JM\right)+\Bcl^2\left(\JME\right)+2\,\Bcl\left(\JM\right)\cdot\Bcl\left(\JME\right)\right].\label{eqn:EF}
\end{equation}
\end{subequations}
Writing this `potential energy' and `free energy' explicitly in
terms of magnetic fields with their origins makes it manifestly
clear that `potential energy' involves physical currents $\JM$ in
$\Vol$ (see Figure~\ref{fig:Decomp}) and the `free energy' involves
physical currents $\JME$ in the external universe (see
Figure~\ref{fig:Closed}). Generally some magnetic energy must be
pilfered from currents $\JME$ in the external universe if $\Bcl$ is
completely dissipated or converted to kinetic energy in a solar
eruption and some magnetic energy must be pilfered from the external
universe to replace the flux threading the boundary that is produced
by coronal currents $\JM$ (see Figure~\ref{fig:BP:Internal}). This
thievery makes `free energy' a dubious concept.\par
The minimum energy state $\Energy_{\mathrm{Potential}}$ is achieved \emph{if
and only if all of the current sources of that potential field are external
to the volume} in $\VolC$, which in terms of Equation~(\ref{eqn:B:Energy:All})
is:
\begin{subequations}
\begin{align}
  \BPot\left(\JM\right)=&0\qquad&\x\in\Vol,\label{eqn:Energy:P}\\
  \BPot=&\BPot\left(\JME\right)\qquad&\x\in\Vol,\\
  \Bcl\left(\JM\right)=&0\qquad&\x\in\Vol,\label{eqn:Energy:Bcl}\\
  \Bcl\left(\JME\right)=&0\qquad&\x\in\Vol,
\end{align}
\end{subequations}
i.e., $\B=\BPot\left(\JME\right)=-\grad\psi\left(\JME\right)$ for
$\x\in\Vol$.  This critical caveat elucidates important assumptions
underlying the achievability of this minimum energy state for the post
eruptive state of the corona. Note that if~(\ref{eqn:Energy:Bcl}) is
true then~(\ref{eqn:Energy:P}) must be true\----there must be a magnetic
field component that closes in $\Vol$ for there to be a potential
field $\BPot\left(\JM\right)$ produced by currents $\JM$ in
$\Vol$. However,~(\ref{eqn:Energy:P}) may be true
when~(\ref{eqn:Energy:Bcl}) is false, e.g., the case of (core)
currents $\JM_\mathrm{C}$ sheathed by the opposing (neutralizing)
currents $\JM_\mathrm{S}$ that shield the boundary $\Surf$ from flux
produced by any internal currents $\JM=\JM_\mathrm{C}+\JM_\mathrm{S}$
where
$\nhat\cdot\bBS\left(\JM_\mathrm{C}+\JM_\mathrm{S};t,\x\right)|_\Surf=0$.\par
The traditional minimum energy proof \cite[]{Priest2014} leaves the
reader with the impression that $\BPot$ and $\Bcl$ are
independent. However this impression is destroyed by
Equations~(\ref{eqn:EF})\--(\ref{eqn:EP}) which include the origin of these
fields.  These fields are only independent when there is no flux threading the
boundary produced by \emph{internal} currents $\JM$ in $\Vol$, i.e., all
internal currents are perfectly shielded, and additionally no currents thread
the boundary, $\nhat\cdot\JM|_\Surf=0$,
\begin{subequations}
\begin{align}
  \BPot\left(\JM\right)=&0\qquad&\x\in\Vol,\label{eqn:Energy:P1}\\
  \BPot=&\BPot\left(\JME\right)\qquad&\x\in\Vol,\\
  \Bcl=&\Bcl\left(\JM\right)\qquad&\x\in\Vol,\\
  \Bcl\left(\JME\right)=&0\qquad&\x\in\Vol.
\end{align}
\end{subequations}
In this case it is possible to dissipate all the closed field, $\Bcl$ and hold
$\nhat\cdot\B|_\Surf$ constant on the boundary without modifying the external
universe $\JME$.  However, $\BPot\left(\JM\right)=0$ for $\x\in\Vol$
does not, in general, hold for the pre-eruptive state of the solar corona
\cite[]{Schuck2022}.\par
Suppose instead that the volume of interest $\Vol$ starts with the initial
magnetic field described by~(\ref{eqn:B:Energy:All}) with the general current
systems $\JMEI$ and $\JMI$. The potential field of the initial state is then
\begin{equation}
\BPot=\BPot\left(\JMEI\right)+\BPot\left(\JMI\right)\qquad\x\in\Vol.
\end{equation}  
The important question for a coronal volume $\Vol$ is not whether a potential
field is the minimum energy state of that volume (it is!), but rather whether
that state is \emph{accessible} from an initial state in $\Vol$ by \emph{only}
dissipating energy in the volume (\Referee{it possible but unlikely!}). Possible examples of
a rapid dissipation process where this question arises involve solar campfires
\cite[]{Berghmans2021}, jets \cite[]{Newton1942}, flares
\cite[]{Carrington1859}, and coronal mass ejections \cite[]{Tousey1973a}.  If
we then hold $\BPot$ constant on the boundary (photosphere) and \emph{only}
rapidly dissipate currents in $\Vol$ the volume of interest (the corona) then
\begin{enumerate*}
\item All currents through $\Surf$ must quickly rearrange so that they
  close in $\VolC$ (the convection zone) which leads to
  $\Bcl\left(\JMEF\right)=0$ for $\x\in\Vol$, where $\JMEF$ and $\JMF$
  are the current systems in the final state.
\item There can be no currents $\JMF$ in $\Vol$, i.e., $\Bcl\left(\JMF\right)=0$
  which then implies $\BPot\left(\JMF\right)=0$.
\item The convection zone currents must rapidly rearrange to replace
  the flux threading the boundary that was initially produced by
  currents in the coronal volume:
\end{enumerate*}
  \begin{equation}
\BPot\left(\JMEF\right)=\BPot\left(\JMEI\right)+\BPot\left(\JMI\right)\qquad\x\in\Vol.   
  \end{equation}
This scenario where the convection zone responds nearly instantaneously to the
dissipation of currents in the corona as it relaxes to a current free
potential state is at odds with the high Alfv{\'e}n speeds in the corona
and low Alfv{\'e}n speeds in the convection zone. Furthermore, this
requires new currents $\JMEF$ in the convection zone to replace the flux
threading the photosphere that was initially produced by coronal currents
$\JMI$. In other words, the convection zone must \emph{add energy} to $\Vol$
during the eruption for the solar atmosphere to achieve a potential state
consistent with the initial boundary condition. In this scenario the
traditional `free energy' is not really free, nor is the energy necessary to
reach the potential state completely contained in the corona prior to the
eruption. As such the `free energy' calculation is \Referee{dubious} for this scenario.\par
A more likely scenario is that currents through the boundary
(photosphere) change, but more importantly, coronal currents rearrange into
thin chromospheric current layers $\KCF$ to minimize their energy and shield
the photosphere from changes in coronal currents \cite[see the magnetic analog
  to Thomson's theorem derived in][]{Fiolhais2008}. The post-eruptive
  state of the solar atmosphere above the photosphere is then
\begin{subequations}
\begin{equation}
  \B=\BPot\left(\JMEF\right)+\Bcl\left(\JMEF\right)+\BPot\left(\JMF+\KCF\right)+\Bcl\left(\JMF+\KCF\right)\qquad\x\in\Vol.   
\end{equation}
The potential field as inferred from the photosphere will remain constant and
it continues to be physically produced both by external $\JMEF$ and internal
$\JMF+\KCF$ currents, but it is primarily the corona/chromosphere
responding to changes in coronal currents not the convection zone
\begin{equation}
\B\neq\BPot=\BPot\left(\JMEF\right)+\BPot\left(\JMF+\KCF\right)=\BPot\left(\JMEI\right)+\BPot\left(\JMI\right)\qquad\x\in\Vol,
\end{equation}
with $\BPot\left(\JMEF\right)\approx\BPot\left(\JMEI\right)$ and
$\BPot\left(\JMF+\KCF\right)\approx\BPot\left(\JMI\right)$. The
corona\--chromosphere system above the photosphere will be
non-potential post-eruption
\begin{equation}
\grad\cross\bBS\left(\JMF+\KCF;t,\x\right)\neq0\qquad\x\in\Vol.  
\end{equation}
\end{subequations}
Thus, the solar atmosphere cannot achieve the minimum energy state and the
`free energy' calculation is \Referee{dubious} for this scenario as well.\par 
\Referee{The scenario where the free energy is most relevent corresponds to the limit between the two scenarios above, when all the currents in the volume $\Vol$ are pushed to the boundary $\Surf$ in the form of current sheets, i.e., in the photosphere or at infinity $|\x|\rightarrow\infty$. This is the magnetic analog of Thomson's theorem \cite[]{Fiolhais2008} which preserves $\nhat\cdot\B|_\Surf$ and establishes a potential field in $\Vol$.
Then the energy that may be released in an eruption through dynamics and heating is exactly the free energy. Of course, this scenario will generate large forces in the photosphere, and in particular torsional forces which cannot be balanced by pressure gradient forces. These forces should manifest themselves as observable changes in the plasma flows and horizontal magnetic fields.}\par
\subsection{The Evolution of the Magnetic Potential Energy With Attribution}
Using identities~(\ref{eqn:Div_psi_f}) and~(\ref{eqn:Gauss}) on
the first row of Equation~(\ref{eqn:EMF}), the magnetic energy in $\Vol$ becomes
\begin{align}
  \Energy=  &\overbrace{\frac{1}{8\,\pi}\left[
      \oint\limits_\Surf{dS}\,\MagPotential\left(\JM\right)\,\nhat\cdot\BPot\left(\JM\right)
     +\oint\limits_\Surf{dS}\,\MagPotential\left(\JME\right)\,\nhat\cdot\BPot\left(\JME\right)
 +2\,\oint\limits_\Surf{dS}\,\MagPotential\left(\JME\right)\,\nhat\cdot\BPot\left(\JM\right)\right]}^{\Energy_\mathrm{Potential}=\frac{1}{8\,\pi}\,\oint\limits_\Surf{dS}\,\MagPotential\,\nhat\cdot\BPot}\nonumber\\
&\qquad+\underbrace{\frac{1}{8\,\pi}\int\limits_\Vol{d^3x}\,\left[\Bcl^2\left(\JM\right)+\Bcl^2\left(\JME\right)+2\,\Bcl\left(\JM\right)\cdot\Bcl\left(\JME\right)\right]}_{\Energy_\mathrm{Free}}.\label{eqn:EM}
\end{align}
The volume integral computation requires modeling, \MHD{} simulations, or
very dense coronal magnetic field observations to calculate the integrals
involving magnetic fields that close in $\Vol$.  However, the surface
integrals may be computed from boundary observations alone.  Using
$\BPot\left(\JM\right)=\BPot-\BPot\left(\JME\right)$ and
$\MagPotential\left(\JM\right)=\MagPotential-\MagPotential\left(\JME\right)$,
each surface term may be computed independently as each surface integral is invariant in isolation under the local
gauge transformation $\MagPotential\rightarrow\MagPotential+\MagPotential_0$
where $\MagPotential_0$ is a constant
\begin{subequations}
\begin{equation}
\oint\limits_\Surf{dS}\,\MagPotential\,\nhat\cdot\BPot=\oint\limits_\Surf{dS}\,\left(\MagPotential+\MagPotential_0\right)\,\nhat\cdot\BPot,   
\end{equation}  
because of the solenoidal property of magnetic fields
\begin{equation}
\oint\limits_\Surf{dS}\,\MagPotential_0\,\nhat\cdot\BPot=\MagPotential_0\,\oint\limits_\Surf{dS}\,\nhat\cdot\BPot=0.
\end{equation}
\end{subequations}
\par 
Consider Equation~(\ref{eqn:EM}) in the solar context where $\Vol$ represents
the volume from the photosphere up through the corona and $\VolC$ represents
the convection zone below the photosphere. Then $\BPot$ represents the
traditional potential field computed from the normal component of the magnetic
field in the photosphere that satisfies $\BPot\rightarrow0$ as
$\left|\x\right|\rightarrow\infty$. Furthermore, the three surface integrals
in~(\ref{eqn:EM}) may be computed from photospheric vector magnetograms with
\textsf{C}arl's \textsf{I}ndirect \textsf{C}oronal \textsf{C}urrent
\textsf{I}mager (\CICCI{}) described in \cite{Schuck2022} which computes the
surface values of both $\BPot$ and
$\BPot\left(\JME\right)$.\footnote{$\BPot\left(\JME\right)=\BLT$ but
$\BPot\left(\JM\right)\neq\BGT$ in the notation of \cite{Schuck2022}.} The
\CICCI{} software is released at the project \textsf{gitlab}
(\url{https://git.smce.nasa.gov/cicci}) under a NASA open source license. The
sum of the three surface integrals is simply the potential field energy in
$\Vol$. If this sum changes during eruptive phenomena then the traditional
potential field, $\BPot$, has changed. In principle $\BPot$ (and $\psi$) may
remain constant if changes in $\nhat\cdot\BPot\left(\JME\right)$ and
$\nhat\cdot\BPot\left(\JM\right)$ cancel out or changes in
  $\nhat\cdot\BPot\left(\JME\right)$ and $\nhat\cdot\BPot\left(\JM\right)$ are
  balanced overall by the changes in mutual energy\----changes in the angle
  between $\BPot\left(\JME\right)$ and $\BPot\left(\JM\right)$ in $\Vol$. 
However, the former cancellation requires detailed balance between
changes in coronal $\JM$ and convection zone $\JME$ currents\----collusion
between $\Vol$ and $\VolC$!  The three individual surface integrals may be
tracked in observations and simulations to determine how the \emph{origin} of
the flux threading the photosphere changes during eruptions and how a detailed
balance is maintained if the photospheric flux remains constant during
explosive coronal phenomena. These surface terms provide a definitive test: is
the convection zone responding to replace flux lost during the eruption or is
the corona/chromosphere system responding to shield the photosphere from
losing flux during the eruption?\par

\section{Summary and Conclusions: Implications for Modeling and Observation\label{sec:Conclusions}}
This work has described the attribution of magnetic fields to current systems
for astrophysical problems. A common approach in solar physics is to decompose
a general magnetic field in $\Vol$ into a potential field $\BPot$ that threads
the boundary $\Surf$ with flux and a component $\Bcl$ that closes on itself
within the volume $\Vol$. Both of these components can have their physical
origin in currents $\JM$ in the internal volume $\Vol$ and $\JME$ in the
external universe $\VolC$. Thus, this representation
$\left(\BPot,\Bcl\right)$, while mathematically convenient, entangles
magnetic field that has its physical origin inside the volume of interest
$\Vol$ with magnetic field that has its physical origin outside the volume
of interest in the external universe $\VolC$. In particular, the naive
implementation of the potential magnetic field $\BPot$ creates a cognitive
dissonance that $\BPot$ is potential and curl free $\grad\cross\BPot=0$ for
$\x\in\Vol$ but physically generated by currents $\JM$ in $\Vol$ (see
Figure~\ref{fig:Decomp} and discussion in \S\ref{sec:Intro},
Introduction). Alternatively, there can be magnetic field in $\Vol$ produced
by currents $\JME$ in $\VolC$ and magnetic field in $\VolC$ produced by
currents $\JM$ in $\Vol$ when no \emph{net} magnetic flux threads $\Surf$,
the boundary between $\Vol$ and $\VolC$ (see Figure~\ref{fig:Closed} and
discussion in \S\ref{sec:Helicity}). We have described how these non
sequiturs may be resolved by attributing the magnetic field to its origin
in $\JM$ or $\JME$ first and then decomposing these fields into potential
$\BPot\left(\JM\right)$ and $\BPot\left(\JME\right)$ and closed
$\Bcl\left(\JM\right)$ and $\Bcl\left(\JME\right)$ components. \par
As presented in \S\ref{sec:Helmholtz}, the computation of the magnetic
field produced by known internal and unknown external current sources,
$\bBS\left(\JM;t,\x\right)$ and $\bBS\left(\JME;t,\x\right)$,
respectively, requires the computation of a Biot-Savart integral for
$\bBS\left(\JM;t,\x\right)$ which establishes cause and effect between
the internal current $\JM$ and the corresponding magnetic field. This
presentation intentionally emphasized this fundamental and intuitive
formulation.  However, Biot-Savart integrals presented to compute
$\AV\left(\JM;t,\x\right)$ and the corresponding field
$\bBS\left(\JM;t,\x\right)$ are computationally intensive for
sources $\JM$ in $\Vol$ and often cannot be performed for
$\AV\left(\JME;t,\x\right)$ and $\bBS\left(\JME;t,\x\right)$ because
$\JME$ is not known in $\VolC$.  From a practical perspective,
constructing the magnetic field produced by external current sources
$\bBS\left(\JME;t,\x\right)$ via the Helmholtz decomposition first will
be more computationally efficient (see \S\ref{sec:Helmholtz}). This is
particularly advantageous when only the magnetic energy is of
interest, i.e., when the vector potentials $\AV\left(\JM;t,\x\right)$
and $\AV\left(\JME;t,\x\right)$ are not needed. This approach involves
the evaluation of only surface integrals instead of convolutions over
the entire volume of interest $\Vol$. Once $\bBS\left(\JME;t,\x\right)$
has been computed, the magnetic field produced by internal currents
may then be constructed by subtraction
$\bBS\left(\JM;t,\x\right)=\B-\bBS\left(\JME;t,\x\right)$, thereby
establishing attribution of the magnetic field to a current system in a
particular domain, i.e., $\JM$  in $\Vol$ or $\JME$ in
$\VolC$, respectively.  The potential and closed components of
these magnetic fields may then be constructed by standard methods (see
\S\ref{sec:Relative} and \S\ref{sec:Attribution}). The last
computationally intensive piece is to construct the closed vector
potentials $\Acl\left(\JM;t,\x\right)$ and
$\Acl\left(\JME;t,\x\right)$. We provide direct approaches in
\S\ref{sec:Attribution} and some additional approaches for the latter
vector potential in Appendix \ref{sec:AV:JME}. Note that the general
results of this work are gauge agnostic, and so our results are not
tied in any way to particular computational approaches or choices of
gauge.\par
Previous work demonstrated that the relative magnetic helicity in
  Equation~(\ref{eqn:HR}) from \citeauthor{Berger1984a} and \cite{Finn1985}
  may be decomposed into the gauge invariant `self' and `mutual' helicities in
  Equation~(\ref{eqn:Berger2003}) from \cite{Berger1999,Berger2003}. This
  decomposition uses the terms `self' to describe the linking of closed field
  $\Bcl$ with itself, $\Acl$, in $\Vol$ and `mutual' to describe the linking
  of closed field $\Bcl$ with the open field $\APot$ in
  $\Vol$. \cite{Longcope2008} point out that these definitions of `self' and
  `mutual' helicity are conceptually distinct from the `self' and `mutual'
  helicity of \emph{isolated} flux tubes where the `self' helicity depends on
  the internal field of each isolated flux tube and the `mutual' helicity
  describes how pairs of tubes are interlinked. \cite{Longcope2008} develop
  further definitions of ``unconfined self helicity'' and ``additive self
  helicity'' based on relative magnetic helicity in sub-volumes of $\Vol$
  \cite[See also][]{Malanushenko2009,Valori2020}.\par
 The novel magnetic field decompositions described in this paper produce
 natural extensions of relative magnetic helicity and magnetic energy that
 incorporate the origin of the magnetic fields in currents $\JM$ in the domain
 of interest $\Vol$ or $\JME$ in the external domain $\VolC$ beyond the
 boundary $\Surf$.  As such, we propose new conceptual definitions of self and
 mutual helicity in $\Vol$ that are attributed to their current sources $\JM$
 and $\JME$ in $\Vol$ and $\VolC$ respectively.  We have extended
 \citeauthor{Berger1998}'s (\citeyear{Berger1998,Berger2003}) representation
 to eight gauge invariant terms that simultaneously describe the origin of the
 magnetic fields in $\JM$ and $\JME$ and how the field components defined by
 $\Surf$, e.g., $\left(\Acl,\Bcl\right)$ and $\left(\APot,\Bcl\right)$, link
 in relative magnetic helicity. \Referee{Seven of these terms are independent.}
 The sum of \Referee{the} eight terms recovers previous results in
 Equation~(\ref{eqn:Berger2003}). Combinations of these terms motivate the new
 definitions of self helicity and mutual helicity: \begin{enumerate*} \item
   internal relative helicity $\Helicity\left(\JM,\JM\right)$ \---- the self
   helicity in $\Vol$ of magnetic fields produced by currents $\JM$ in
   $\Vol$; \item external relative helicity $\Helicity\left(\JME,\JME\right)$
   \---- the self helicity in $\Vol$ of magnetic fields produced by currents
   $\JME$ in $\VolC$;
     \item internal-external relative helicity
       $\Helicity\left(\JM,\JME\right)$+$\Helicity\left(\JME,\JM\right)$ \----
       the mutual helicity in $\Vol$ of magnetic fields produced by currents
       $\JM$ in $\Vol$ with magnetic fields produced by currents $\JME$ in
       $\VolC$.\end{enumerate*} Tracking the evolution of  \Referee{the seven
   independent} terms will provide insight into how magnetic linkages change
 during fundamental stellar processes such as flux emergence, coronal heating,
 and eruptive phenomena. However, tracking the evolution of these \RefereeR{eight} terms
 requires access to dense magnetic field measurement presently only available
 in simulations and modeling \cite[]{Pariat2017}. Nonetheless, further
 consideration of the helicity transport across boundaries in terms of this
 framework may reveal observables that can be computed from photospheric
 observations alone \cite[see the helicity transport representation developed
   in][]{Schuck2019}. We have also decomposed the magnetic energy in a volume
 into terms that describe the origin of the magnetic fields. This
 representation results directly in terms that may be computed from surface
 observations alone, and when combined with new theoretical and computational
 techniques \cite[]{Schuck2022}, it has the potential to reveal the interplay
 between the photosphere, convection zone, and corona during solar eruptive
 phenomena.\par
The concept of cause and effect from currents to magnetic fields outlined in
this work has broad application to solar physics. Attributing changes in
current systems that lead to changes in magnetic structure has the potential
to reveal causality in `sympathetic' solar eruptions
\cite[]{Bumba1993}. Furthermore, combining the attribution of currents in
simulations presented here with new attribution techniques, such as \CICCI{},
applicable to the photospheric surface has the potential to unambiguously
connect the photospheric/chromospheric magnetic fingerprints of eruptive
phenomena to coronal current systems, e.g., the photospheric fingerprints of
the formation of the flare current sheet in the corona.\par
We often ignore the interaction between the external universe $\VolC$ or
equivalently boundary sources on $\Surf$ and the evolution of magnetic fields
in our volume of interest $\Vol$. However, the current sources in $\VolC$ are
often major players in the evolution of the magnetic field in modeling the
evolution in $\Vol$.  Connecting the magnetic field with its origin in
currents provides a deeper and clearer understanding of the evolution of
astrophysical plasmas and \MHD{} simulations.\par

\newcommand{\MarkAKS}{NNH17ZDA001N-LWS}
\newcommand{\MarkAKP}{17-LWS17_2-0065}
\newcommand{\MarkAKT}{Investigating Magnetic Flux Emergence with Modeling and
  Observations to Understand the Onset of Major Solar Eruptions}

\newcommand{\PeteAKS}{NNH17ZDA001N-LWS}
\newcommand{\PeteAKP}{16-LWS16_2-0065}
\newcommand{\PeteAKT}{Developing Vector Magnetic Maps from SDO/HMI that can Drive Space Weather Models}

\newcommand{\PETEAKS}{NNH21ZDA001N-LWS}
\newcommand{\PETEAKP}{21-LWS21_2-0042}
\newcommand{\PETEAKT}{The Origin of the Photospheric Magnetic Field: Mapping Currents in the Chromosphere and Corona}

\newcommand{\RickAKS}{HISFM18}
\newcommand{\RickAKP}{17-HISFM18-0007}
\newcommand{\RickAKT}{Magnetic Energy Buildup and Explosive Release in the Solar Atmosphere}

\newcommand{\JamesAKS}{NNH16ZDA001N-LWS}
\newcommand{\JamesAKP}{16-LWS16_2-0076 }
\newcommand{\JamesAKT}{Implementing and Evaluating a Vector-Magnetogram-Driven
  Magnetohydrodynamic Model of the Magnetic Field in the Low Solar Atmosphere}

\newcommand{\KalmanAKS}{NNH18ZDA001N-HSR}
\newcommand{\KalmanAKP}{18-SUN18_2-0035}
\newcommand{\KalmanAKT}{Investigating Magnetic Flux Rope Emergence as the 
  Source of Flaring Activity in Delta-Spot Active Regions}

\newcommand{\BarbaraAKS}{??}
\newcommand{\BarbaraAKP}{??}
\newcommand{\BarbaraAKT}{Center for HelioAnalytics (GSFC)}

\newcommand{\PulkkinenAKS}{NNH17ZDA001N-LWS}
\newcommand{\PulkkinenAKP}{}
\newcommand{\PulkkinenAKT}{Physics-based modeling of the magnetosphere-ionosphere-thermosphere-mesosphere system under Carrington-scale solar driving: response modes, missing physics and uncertainty estimates}

\begin{acknowledgments}
The authors thank the referee. The authors recognize useful conversation with
James Leake, Lars Daldorf, Dana Longcope, and Brian Welsch.  Peter W. Schuck
dedicates his work on this paper to Henry J. Schuck.  The authors acknowledge
support from
the NASA Living with a Star (H-LWS) Focused Science Topic programs: \PETEAKS{}
``\PETEAKT{}'' (Schuck, Linton), \PeteAKS{} ``\PeteAKT{}'' (Schuck),
\JamesAKS{} ``\JamesAKT{}'' (Linton, Schuck), and \MarkAKS{} ``\MarkAKT{}''
(Linton, Schuck), \Referee{\PulkkinenAKS{} ``\PulkkinenAKT'' (Schuck)}; the NASA Supporting
Research (H-SR) programs: \KalmanAKS{} ``\KalmanAKT {}'' (Linton); and the
Office of Naval Research (Linton); and from the NASA Internal Science Funding
Model (H-ISFM) program ``\RickAKT{}'' (Schuck).  \par
\end{acknowledgments}

\renewcommand{\theequation}{\theappendix.\arabic{equation}}
\numberwithin{equation}{section}
\appendix
\section{Woltjer's Boundary Condition\label{sec:Boundary}}
The \cite{Woltjer1958} boundary condition for a magnetically closed
system $\Vol$ is $\partial\AV/\partial t|_\Surf=0$.  This boundary condition
certainly preserves the helicity in Equation~(\ref{eqn:Woltjer}), but its
complete physical consequences are not manifest and so we clarify them
below.\par
From Equation~(\ref{eqn:Woltjer}),
the necessary and sufficient condition for helicity invariance in the Gibbs gauge
is:
\begin{equation}
\frac{d H}{d t}
=\oint\limits_\Surf{dS}\,\nhat\cdot\AV\cross\frac{\partial\AV}{\partial t}=0.\label{eqn:Woltjer:S}
\end{equation}
The  incomplete Gibbs gauge is defined by a
transformation from a potential pair $\left(\varphi',\AV'\right)$ to
$\left(0,\AV\right)$ via
\begin{subequations}
\begin{align}
\AV=&\AV'+\grad\Lambda',\label{eqn:Gauge:A}\\
\varphi=&\varphi'-\frac{1}{c}\,\frac{\partial\Lambda'}{\partial
  t}=0,\label{eqn:Gauge:Phi}
\end{align}
where
\begin{equation}
\Lambda'=c\,\int\limits_{-\infty}^t{dt}\,\varphi',
\end{equation}
 \end{subequations}
and
\begin{subequations}
\begin{align}
\E=&-\frac{1}{c}\,\frac{\partial \AV'}{\partial t}-\grad\varphi'=-\frac{1}{c}\,\frac{\partial \AV}{\partial t},\\ 
\B=&\grad\cross\AV'=\grad\cross\AV.
\end{align}
\end{subequations}
Rewriting Equation~(\ref{eqn:Woltjer:S}) in an arbitrary gauge
\begin{equation}
\frac{d H}{d t}
  =\oint\limits_\Surf{dS}\,\nhat\cdot\AV\cross\frac{\partial\AV}{\partial t}+\oint\limits_\Surf{dS}\,\nhat\cdot\AV\cross\frac{\partial\grad\gauge}{\partial t}+\oint\limits_\Surf{dS}\,\nhat\cdot\grad\gauge\cross\frac{\partial\AV}{\partial t}=0,
\end{equation}  
and using the vector identity  
\begin{equation}  
\grad\cross\left(\phi\,\boldsymbol{a}\right)=\grad\phi\cross\boldsymbol{a}+\phi\,\grad\cross\boldsymbol{a}  
\end{equation}  
with Equation~(\ref{eqn:B}) this becomes
\begin{equation}  
\frac{d H}{d t}
  =\oint\limits_\Surf{dS}\,\nhat\cdot\AV\cross\frac{\partial\AV}{\partial t}
  +\oint\limits_\Surf{dS}\,\nhat\cdot\grad\cross\left(\gauge\,\frac{\partial\AV}{\partial t}-\frac{\partial\gauge}{\partial t}\,\AV\right)
  +\oint\limits_\Surf{dS}\,\nhat\cdot\left(\frac{\partial\gauge}{\partial t}\,\B-\gauge\,\frac{\partial\B}{\partial t}\right).
\end{equation}
The second surface integral involving the curl is identically zero and
the third surface integral is zero for an arbitrary gauge
transformation if $\nhat\cdot\B|_\Surf=0$. Thus, the boundary condition
$\nhat\cdot\B|_\Surf=0$ to ensure gauge invariance is an implicit
assumption in Equation~(\ref{eqn:Woltjer}).\par
The well-known jump conditions on the observable electric and
magnetic fields across a boundary $\Surf$ are: \cite[see pp. 19-20 in][]{Jackson1975}
\begin{subequations}
\begin{align}
\Jump{\nhat\cdot\E}=&4\,\pi\,\sigma,\qquad&  
\Jump{\nhat\cross\E}=&0,\label{eqn:nxE}\\
\Jump{\nhat\cdot\B}=&0,\qquad&  
\Jump{\nhat\cross\B}=&\frac{4\,\pi}{c}\,\KC,\label{eqn:nxB}
\end{align}  
where $\sigma$ is the surface charge, $\KC$ is the surface current, and
$\Jump{a}=\left.a\right|_{\Surf^+}-\left.a^*\right|_{\Surf^-}$ is shorthand
for the jump conditions from above the surface in $\Vol$ (denoted with a
superscript ``+'') to below the surface in $\VolC$ (denoted with a superscript
``-'') \cite[see p. 20 in][]{Jackson1975}. The jump condition on the
tangential components of the magnetic field may be recast as a surface
continuity equation \cite[]{Arnoldus2006}
\begin{equation}
\Jump{\nhat\cdot\grad\cross\B}=-\frac{4\,\pi}{c}\,\gradS\cdot\KC.
\end{equation}
\end{subequations}
The kinematic boundary
condition at a fluid-fluid interface is
\begin{equation}
\Jump{\nhat\cdot\vel}=0.\label{eqn:Kinematic}
\end{equation}
The jump conditions on $\E$ and $\B$ rigorously correspond to jump conditions
on the vector potential in the Gibbs gauge\footnote{Note that for
a \cite{Dupin1813} surface  \cite[in particular see A3.24 in][]{Bladel2007}
\begin{displaymath}
\nhat\cross\left(\grad\cross\AV\right)=\gradS{A_n}-\frac{A_1}{R_1}\,\hat{\boldsymbol{e}}_1-\frac{A_2}{R_2}\,\hat{\boldsymbol{e}}_2-\frac{\partial \AV_\surf}{\partial n},  
\end{displaymath}  
where $A_1$ and $A_2$ are the components of $\AV$ in the principle directions
$\hat{\boldsymbol{e}}_1$ and $\hat{\boldsymbol{e}}_2$ and $R_1$ and $R_2$ are
the respective principle curvatures. Since
$A_1$ and $A_2$ are continuous across $\Surf$ they do not appear in boundary
conditions~(\ref{eqn:BC:B:A}) derived from Equation~(\ref{eqn:nxB}).}
\begin{subequations}
\begin{align}
\Jump{\frac{\partial A_n}{\partial t}}=&-4\,\pi\,c\,\sigma,\qquad&  
\Jump{\frac{\partial\AV_\surf}{\partial t}}=&0,\label{eqn:BC:E:A}\\
\Jump{\AV_\surf}=&0,\qquad&
\Jump{\gradS{A_n}-\frac{\partial\AV_\surf}{\partial n}}=&\frac{4\,\pi}{c}\,\KC.\label{eqn:BC:B:A}
\end{align}  
\end{subequations}
The gauge invariance condition~(\ref{eqn:Gauge}) implies that
$\nhat\cdot\B|_\Surf=\nhat\cdot\grad\cross\AV|_\Surf=0$ and thus
$\nhat\cross\AV|_\Surf=\nhat\cross\grad\Lambda$.\footnote{Note that
$\grad\cross\nhat|_\Surf=0$ for a \citeauthor{Dupin1813} surface.} The jump
condition $\Jump{\partial\AV_\surf/{\partial t}}=0$ in the Gibbs
gauge~(\ref{eqn:Weyl}) combined with the kinematic boundary
condition~(\ref{eqn:Kinematic}) and $\nhat\cdot\B|_\Surf=0$ requires either
$\nhat\cdot\vel^*|_{\Surf^-}=\nhat\cdot\vel|_{\Surf^+}=0$ or
$\Jump{\nhat\cross\B}=0$ (no surface current $\KC$). Since $\KC$ is explicitly
considered by \cite{Woltjer1958}, the former condition
$\nhat\cdot\vel|_\Surf=0$ is implied.  This then implies
$\partial\AV_\surf/{\partial t}|_\Surf=0$ \emph{on the boundary} (not to be
confused with $\Jump{\partial\AV_\surf/{\partial t}}=0$), which is sufficient
to ensure that the surface integral in Equation~(\ref{eqn:Woltjer}) is zero,
i.e., that the system is \emph{sufficiently `isolated,'} to preserve
helicity.\footnote{We emphasize that enforcing gauge invariance is not
sufficient for dynamical conservation of helicity $H$. See
footnote~\ref{ft:Noether}.} This condition does not inherently preclude the
existence of a surface charge $\sigma$ or surface current $\KC$ on $\Surf$ in
boundary conditions~(\ref{eqn:BC:E:A})\--(\ref{eqn:BC:B:A}). Indeed, since
$B_n=0$ and $v_n=0$ on $\Surf$, then the electric field is always normal to
the boundary $\E=E_n\,\nhat$ and the Poynting vector
$c\,\E\cross\B/\left(4\,\pi\right)=\left(\vel_\perp\cross\B_\perp\right)\cross\B_\perp/\left(4\,\pi\right)$
is always tangent to the boundary $\Surf$\----no \emph{net} electromagnetic
energy crosses the boundary, but collusion is permitted! Regardless of
the gauge condition $\grad\cdot\AV$, the jump condition on the tangential
components $\Jump{\AV_\surf}=0$ follows by analogy from the jump conditions on
the tangential components of the electric field. Just as $\Jump{\E_\perp}=0$
because $\partial\B_\perp/\partial t$ must be finite in the surface $\Surf$,
so $\Jump{A_\perp}=0$ because $\B_\perp$ must be finite in the surface
$\Surf$.  However, in analogy with the jump conditions on the normal component
of the electric field, the normal component of the vector potential in the
Gibbs gauge may be discontinuous, $\Jump{A_n}\neq0$. Consequently, the jump
conditions for the surface current $\KC$ involve derivatives of all three
components of the vector potential. Of course if $\AV'$ is in the Coloumb
gauge with $\grad\cdot\AV'=0$ then it is apparent that $\Jump{\AV'_n}=0$
\cite[see p. 242 in][]{Griffiths1999}. \cite{Woltjer1958} imposes quite
reasonable, but more stringent, boundary conditions for a magnetically
closed system, namely, $\partial\AV/\partial t|_\Surf=0$ in the Gibbs gauge,
which requires $\sigma=0$ and $\vel\cross\B|_\Surf=0$. Under these conditions
$\Jump{{\partial\AV}/{\partial t}}=0$ and, in the absence of free charge at
the boundary, the Gibbs gauge reduces to the Coulomb gauge on the boundary
with $\Jump{\AV}=0$ with $\Jump{{\partial\AV_\surf}/{\partial
    n}}={4\,\pi}{c}^{-1}\,\KC$.\par

\section{Other expressions for $\Acl\left(\JME;\lowercase{t},\x\right)$ from $\Bcl\left(\JME;\lowercase{t},\x\right)$\label{sec:AV:JME}}
As mentioned in \S~\ref{sec:H:External}, the
expression~(\ref{eqn:Acl:Simple}) for $\Acl\left(\JME;t,\x\right)$ is
conceptually simple, but involves a computationally intensive
convolution integral of $\Bcl\left(\JME;t,\x\right)$ in
$\Vol$. However, alternative representations may be derived from the
\HD{} of $\Bcl\left(\JME;t,\x\right)$ in Equation~(\ref{eqn:B:V}) of
\S\ref{sec:Helmholtz} in terms of the internal vorticity of
$\Bcl\left(\JME;t,\x\right)$ and its tangential components on the
bounding surface $\Surf$
\begin{equation}
\Bcl\left(\JME;t,\x\right)=
\grad\cross\left[\frac{1}{4\,\pi}\,\int\limits_\Vol{d^3x'}\frac{\grad'\cross\Bcl\left(\JME;t,\x'\right)}{\left|\x-\x'\right|}+\frac{1}{4\,\pi}\,\oint\limits_\Surf{dS'}\,\frac{\nhat'\cross\Bcl\left(\JME;t,\x'\right)}{\left|\x-\x'\right|}\right]\qquad\x\in\Vol.
\end{equation}   This implies
\begin{equation}
\Acl\left(\JME;t,\x\right)=\frac{1}{4\,\pi}\,\int\limits_\Vol{d^3x'}\,\frac{\grad'\cross\Bcl\left(\JME;t,\x'\right)}{\left|\x-\x'\right|}+\frac{1}{4\,\pi}\oint\limits_\Surf{dS'}\,\frac{\nhat'\cross\Bcl\left(\JME;t,\x'\right)}{\left|\x-\x'\right|}\qquad\x\in\Vol.\label{eqn:Acl}  
\end{equation}  
However, as written, this also requires a computationally intensive Biot-Savart
type convolution, but over $\grad'\cross\Bcl\left(\JME;t,\x'\right)$. One
alternative is to substitute~(\ref{eqn:Curl:bBS:JME}) into~(\ref{eqn:Acl})
\begin{equation}
\Acl\left(\JME;t,\x\right)=-\frac{1}{4\,\pi\,c}\,\int\limits_\Vol{d^3x'}\,\frac{1}{\left|\x-\x'\right|}\,\grad'\oint\limits_{\Surf}{dS''}\,\frac{\nhat''\cdot\JM\left(t,\x''\right)}{\left|\x'-\x''\right|}+\frac{1}{4\,\pi}\oint\limits_\Surf{dS'}\,\frac{\nhat'\cross\Bcl\left(\JME;t,\x\right)}{\left|\x-\x'\right|}\qquad\x\in\Vol.  
\end{equation}  
Integrating by parts 
\begin{align}
  \Acl\left(\JME;t,\x\right)=&-\frac{1}{4\,\pi\,c}\,\int\limits_\Vol{d^3x'}\,\grad'\left[\frac{1}{\left|\x-\x'\right|}\,\oint\limits_{\Surf}{dS''}\,\frac{\nhat''\cdot\JM\left(t,\x''\right)}{\left|\x'-\x''\right|}\right]+\frac{1}{4\,\pi}\oint\limits_\Surf{dS'}\,\frac{\nhat'\cross\Bcl\left(\JME;t,\x\right)}{\left|\x-\x'\right|}\nonumber\\
  &\qquad-\frac{1}{4\,\pi\,c}\,\grad\int\limits_\Vol\frac{d^3x'}{\left|\x-\x'\right|}\,\oint\limits_{\Surf}{dS''}\,\frac{\nhat''\cdot\JM\left(t,\x''\right)}{\left|\x'-\x''\right|}\qquad\x\in\Vol.  
\end{align}  
Using the Gauss-Ostrogradsky theorem applied to the gradient of a scalar
\begin{equation}
\int\limits_\Vol{d^3x}\,\grad{\phi}=-\oint\limits_\Surf{dS}\,\nhat\,\phi,  
\end{equation}  
$\Acl$ may be written as a double convolution 
\begin{align}
  \Acl\left(\JME;t,\x\right)=&\frac{1}{4\,\pi\,c}\,\oint\limits_\Surf{dS'}\,\oint\limits_{\Surf}{dS''}\,\frac{\nhat'\nhat''\cdot\JM\left(t,\x''\right)}{\left|\x-\x'\right|\,\left|\x'-\x''\right|}+\frac{1}{4\,\pi}\oint\limits_\Surf{dS'}\,\frac{\nhat'\cross\Bcl\left(\JME;t,\x\right)}{\left|\x-\x'\right|}\nonumber\\
  &\qquad-\frac{1}{4\,\pi\,c}\,\grad\int\limits_\Vol\frac{d^3x'}{\left|\x-\x'\right|}\,\oint\limits_{\Surf}{dS''}\,\frac{\nhat''\cdot\JM\left(t,\x''\right)}{\left|\x'-\x''\right|}\qquad\x\in\Vol
\end{align}  
over just boundary values. The last term is simply the gradient of a gauge
function and may be ignored in our gauge invariant approach.\par


\ifx\homepath\overleafhome
\bibliography{bibliography,extra}
\else
\bibliography{bibliography,extra}
\fi

\end{document}